%% file: MCviaQRNs_v3.tex
\documentclass[review]{elsarticle}

\usepackage{xcolor}
\usepackage{url}
\usepackage{graphicx}
 \usepackage{epstopdf}
\usepackage{dsfont}
\usepackage{amssymb,amsmath}
\usepackage{amsthm}
\usepackage[margin=1.3in]{geometry}
\usepackage{amsfonts}
\usepackage{enumerate}
\usepackage[colorlinks = true,
            linkcolor = blue,
            urlcolor  = blue,
            citecolor = blue,
            anchorcolor = blue]{hyperref}
\usepackage{bigints}
\usepackage[multiple]{footmisc}

\usepackage{tikz}
\usetikzlibrary{shapes,arrows}

\input{notation}

\theoremstyle{definition}

\begin{document}
\begin{frontmatter}

\title{Quantum Systems for Monte Carlo Methods\\and Applications to Fractional Stochastic
Processes}


\author{Sebastian F. Tudor}
\address{School of Business, Financial Engineering Division, Hanlon Financial Systems Center,\\Stevens Institute of Technology, Hoboken, NJ 07030, USA}

\author{Rupak Chatterjee}
\address{Department of Physics, Center for Quantum Science and Engineering,\\Hanlon Financial Systems Center, Stevens Institute of Technology, Hoboken, NJ 07030, USA}

\author{Lac Nguyen, Yuping Huang}
\address{Department of Physics, Center for Quantum Science and Engineering,\\Stevens Institute of Technology, Hoboken, NJ 07030, USA}

\begin{abstract}
Random numbers are a fundamental and useful resource in science and engineering with
important applications in simulation, machine learning and cyber-security. Quantum
systems can produce true random numbers because of the inherent randomness at the core
of quantum mechanics. As a consequence, quantum random number generators are an efficient method
to generate random numbers on a large scale. We study in this paper the applications of a viable source of unbiased quantum random numbers (QRNs) whose statistical properties can be arbitrarily programmed without the need for any post-processing and that pass all standard randomness tests of the NIST and Dieharder test suites without any randomness extraction. Our method is based on measuring the arrival time of single photons in shaped temporal modes that are tailored with an electro-optical
modulator. The advantages of our QRNs are shown via two applications: simulation of a fractional Brownian motion, which is a non-Markovian process, and option pricing under the fractional SABR model where the stochastic volatility process is assumed to be driven by a fractional Brownian motion. The results indicate that using the same number of random units, our QRNs achieve greater accuracy than those produced by standard pseudo-random number generators. Moreover, we demonstrate the advantages of our method via an increase in computational speed, efficiency, and convergence.
\end{abstract}


\end{frontmatter}

\section{Introduction}

The motivation for creating an unbiased quantum random number generator is in its wide range of applications in not only secure communication and cryptography, but also in its widespread use in Monte-Carlo (MC) simulations. It is well known that pseudo-random number (PRN) generators are not random at all such that seed-based generators can be used to generate identical strings of random numbers based on the same seed and similar computational systems. Furthermore, it has been shown that the entropy rate also approaches zero when viable random seeds are used up \cite{herrero2017quantum}. In contrast, quantum random numbers (QRN) are physically created by quantum processes that are inherently stochastic. 
That is, one can by no means can predict a sequence of numbers before they are acquired from the source, even if one has full knowledge about the state of the system. In recent years, many methods have been proposed for QRNs producing a uniform distribution, see e.g. \cite{ma2016quantum}, \cite{zhang2017quantum}. Compared to other methods of implementation, which focus on high data rate and test suite scores, our approach is notable where we also demonstrate that QRNs can be generated directly in arbitrary probability distributions, with high-dimensionality, and free of post-processing. It is based on the random arrival time measurement of single photons that have temporal wave-forms shaped by an electro-optical modulator \cite{nguyen2018programmable}. A key application area of such random numbers is for MC simulations for stochastic processes. 


An important task in financial and risk engineering is to study stochastic processes \cite{glasserman2013monte}. The wide spread use of Monte Carlo simulations of these stochastic processes in enterprise level risk management has led to an ever insatiable need for computing power and speed. The creation of these simulation paths using true random
number generators is precisely where these quantum random numbers can play a critical role.

Simulation of stochastic processes in general, and of {\it fractional} stochastic processes in particular, is an important problem in engineering and science. A fractional Brownian motion offers a convenient modeling approach for non-stationary and non-Markovian stochastic processes with statistical self-similarity, and makes natural the use of wavelets for both its analysis and synthesis \cite{flandrin1992wavelet}. Fractal feature analysis and classification in medical imagining can be studied by using a fractional Brownian motion, since the fractal dimension in medical images may be obtained from the Hurst exponent \cite{chen1989fractal}. Furthermore, traffic phenomena is connectionless networks can be modeled and analyzed via a fractional Brownian motion \cite{norros1995use}.

Having the motivation that simulation of fractional stochastic processes and MC techniques are widely applied in science, engineering, and finance, we present in this paper the advantages of using a QRN generator versus its classical counterpart (pseudo-random numbers) for increased computational efficiency and accuracy. Our purpose is to simulate faster and more efficiently stochastic processes that are not analytically tractable. 

We begin with a short description of our QRN generator and show the advances of our experimental device as it compares with similar products available in the industry and other research laboratories. We discuss in Section 3 the fractional Brownian motion (fBM) and its properties, present some well-known simulation methods, and compare the outcomes of the simulation via PRNs versus QRNs in terms of speed and performance. Thereafter, we present in Section 4 some relevant applications to fractional stochastic processes. We first discuss the Rough Fractional Stochastic Volatility (RFSV) model from a seminal paper \cite{gatheral2018volatility} and show the applicability of fractional stochastic processes in volatility modeling in finance. In a similar manner, we define a fractional SABR (fSABR) model and price some path dependent/exotic volatility options under the fSABR model. Finally, we use our QRNs to price these same financial instruments under the fSABR model and compare them to pseudo random number methods. The paper ends with several conclusions and directions for future research.


\section{The Quantum Random Number Generator}



Random numbers are key for a wide range of applications, spanning issues like encryption, cryptography, and numerical simulations in physics, biochemistry, molecular biology or finance, see \cite{kurtsiefer2002quantum}, \cite{acin2012randomness}. Classical approaches for generating random numbers (the so-called pseudo-random numbers, which are generated by complex yet deterministic numerical algorithms on classical/non-quantum computers) are often insufficient to meet the growing demands in data security and accuracy, see for instance \cite{click2011quality}.

\subsection{Experimental setup}

\begin{figure}
\centering
\includegraphics[scale = 0.44]{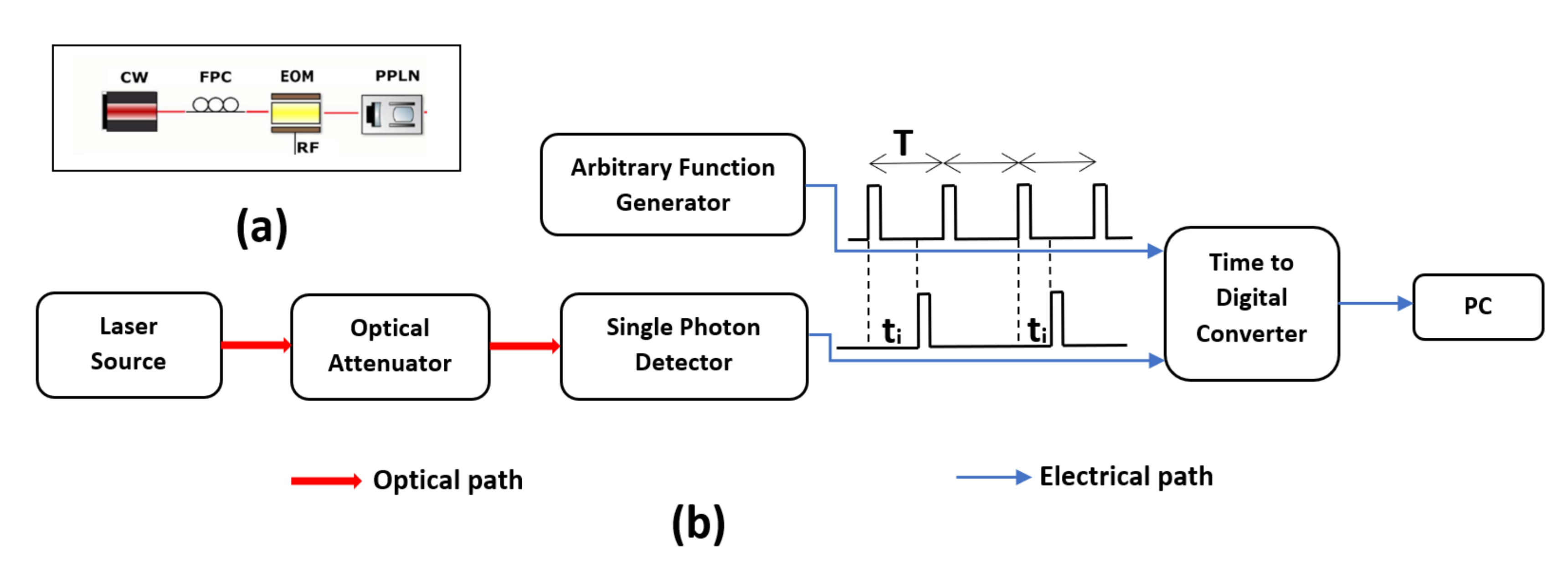}
\caption{(a) The laser source; (b) diagram of the experimental setup for the QRN Generator}
\label{expSet} 
\end{figure}

In a recent paper \cite{nguyen2018programmable}, we present a viable source of unbiased quantum random numbers whose statistical properties can be programmed without need for post-processing. The device is based on measuring the arrival times of single photons in shaped temporal modes, as illustrated in Figure \ref{expSet}. We also demonstrate direct generation of genuine QRNs in user-defined, arbitrary distribution functions, with superior statistical properties than existing devices.

\subsection{NIST and Dieharder tests}

\begin{figure}
\centering
\includegraphics[scale = 0.43]{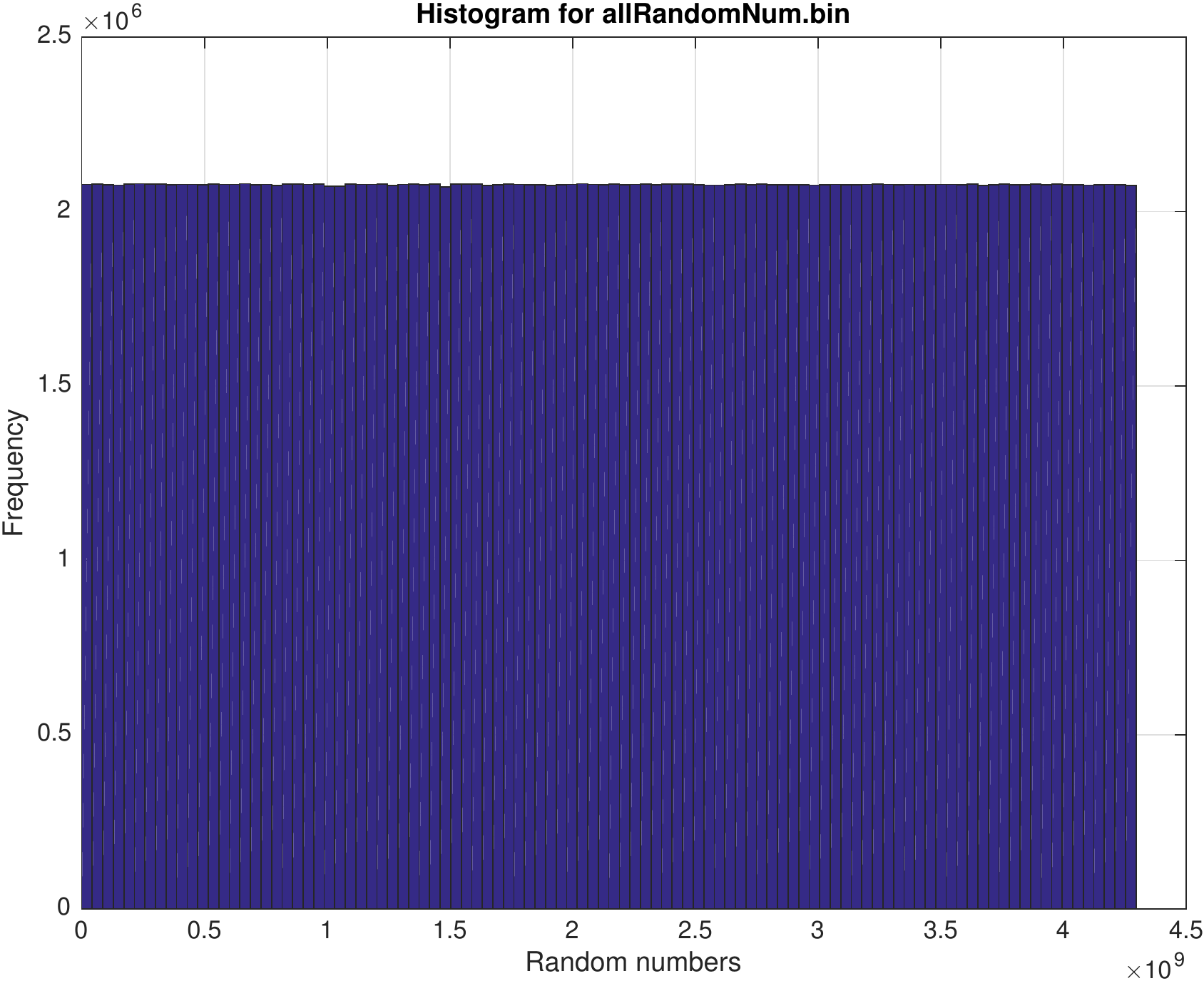}
\includegraphics[scale = 0.43]{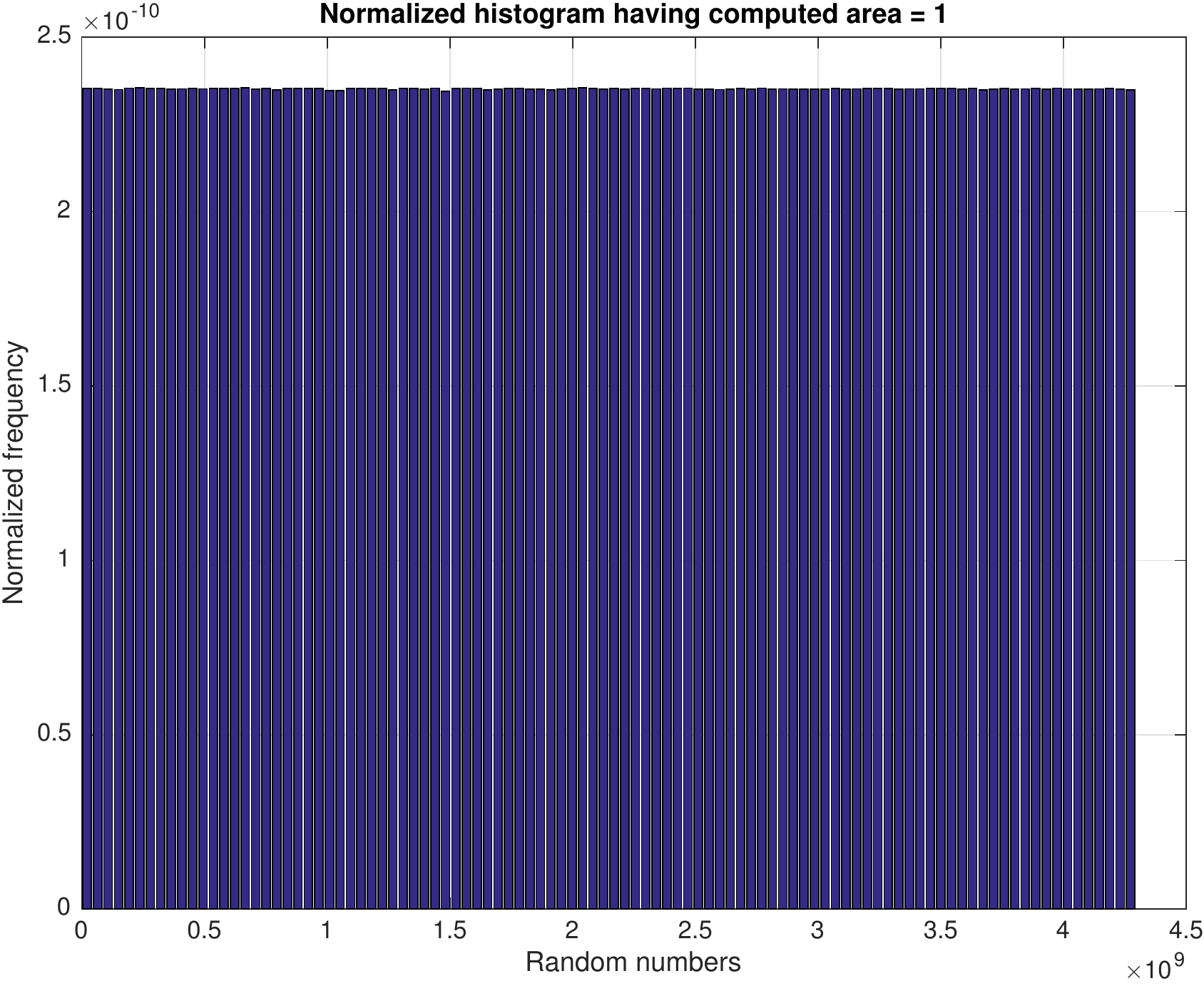}
\caption{The histogram and the normalized histogram for allRandomNum.bin}
\label{hist}

\
\\

\
\\

\includegraphics[scale = 0.43]{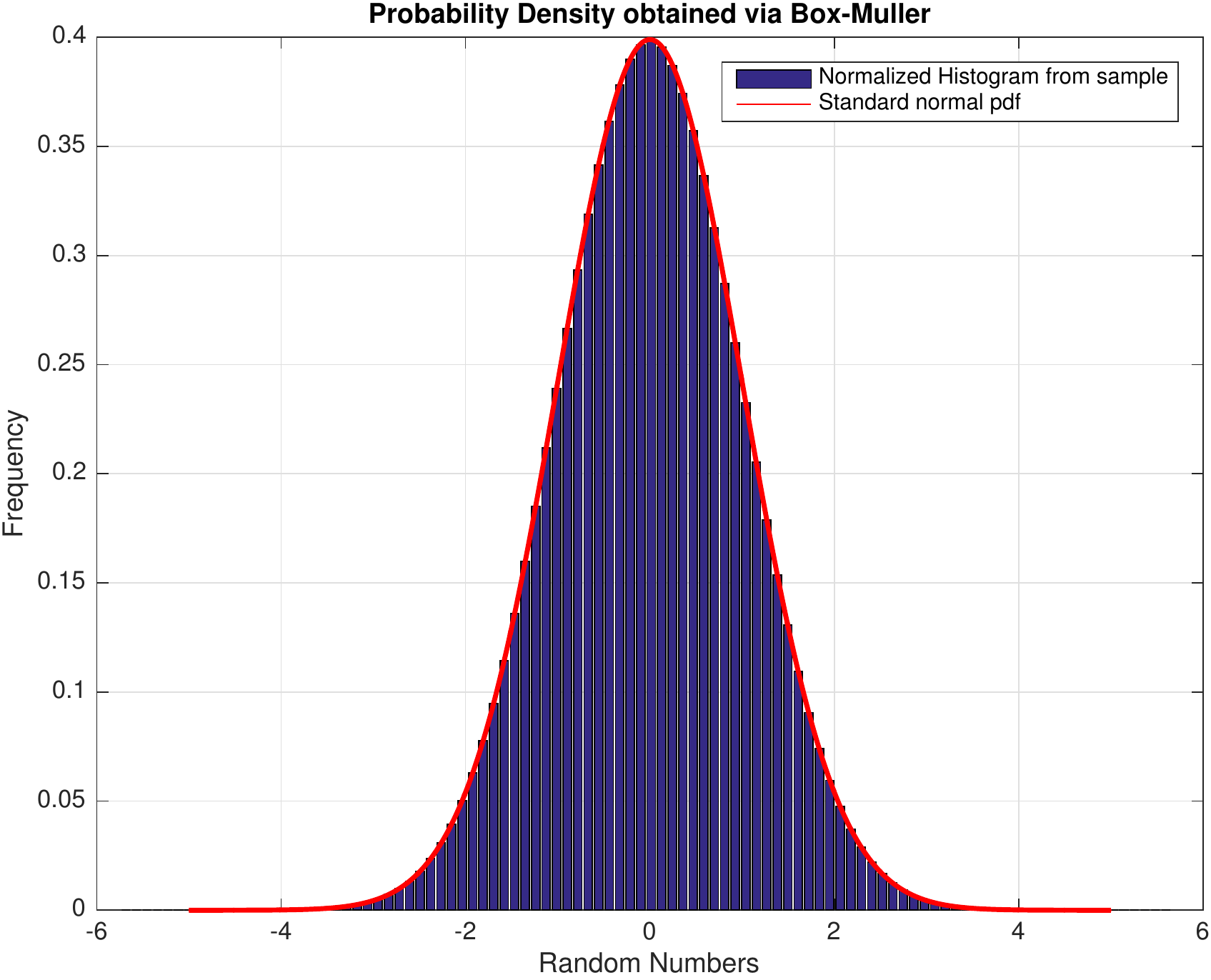}
\includegraphics[scale = 0.43]{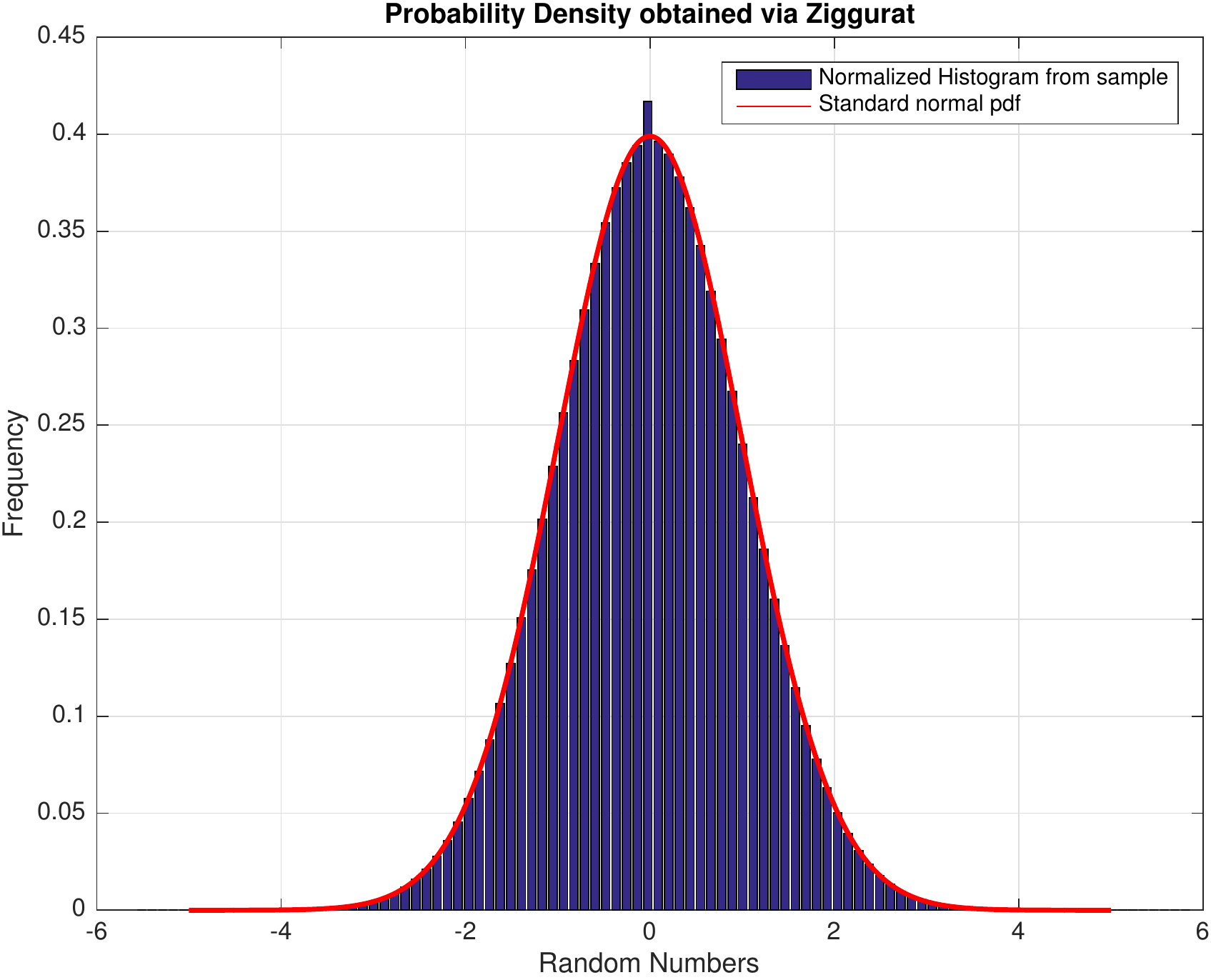}
\caption{Probability density functions obtained from a sample of $2\x10^8$ uniformly distributed random integers via Box--Muller and Ziggurat transform, respectively. The algorithms were implemented in C++ on a 2.4 GHz 6-Core Intel Xeon Processor. The execution time is 173.011 and 142.731 seconds, respectively.}
\label{hist2}
\end{figure}

To assess the quality of our QRN generator, we first generate a large amount of random numbers. The data (QRNs) contains $2.07732758\x 10^8$ random integers in the set $$n_i\in[0,4.294967288\x10^9]\cap\Z,\ i=1,2,\dots,207732758$$ written on 32 bits. The histogram and the normalized histogram of the sample are given in Figure \ref{hist}. Note that the generated numbers are uniformly distributed. By applying the Ziggurat algorithm or the Box--Muller transform, one can transform the uniform distribution into a standard normal distribution. The summary of results is given in Figure \ref{hist2}. In particular, generating \emph{normally} distributed samples is straightforward from the experimental point of view which makes preprocessing/transformation of our data unnecessary.

\begin{figure}
\centering
\includegraphics[scale = 0.8]{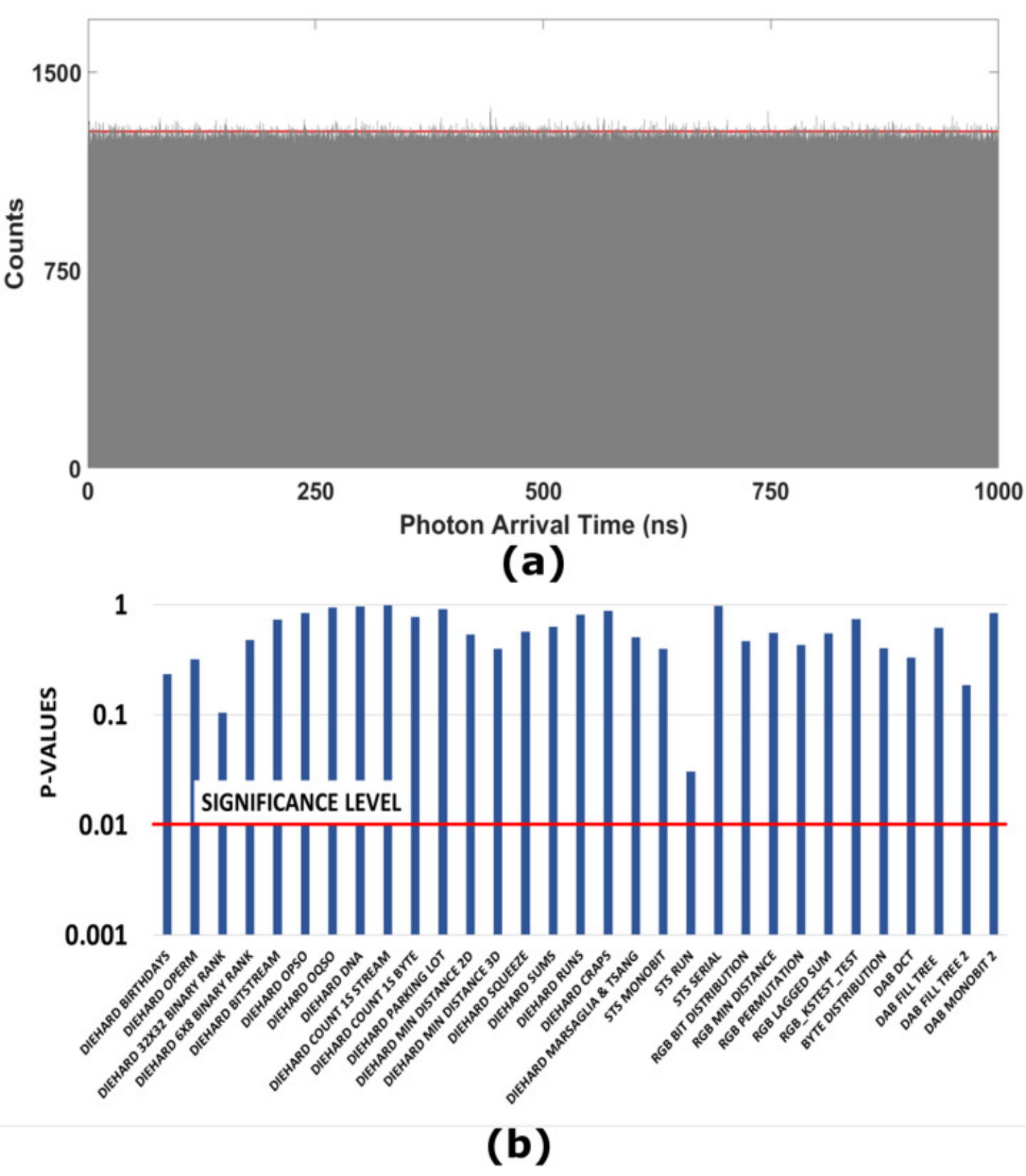}
\caption{(a) Histogram of photon arrival times; (b) P-values for Dieharder test routines. Note that our QRNs pass all the statistical tests from the Dieharder test suite. This figure is a reproduction of Figure 2 in \cite{nguyen2018programmable}.}
\label{perfQRN}
\end{figure}

\begin{figure}
\centering
\includegraphics[scale=0.8]{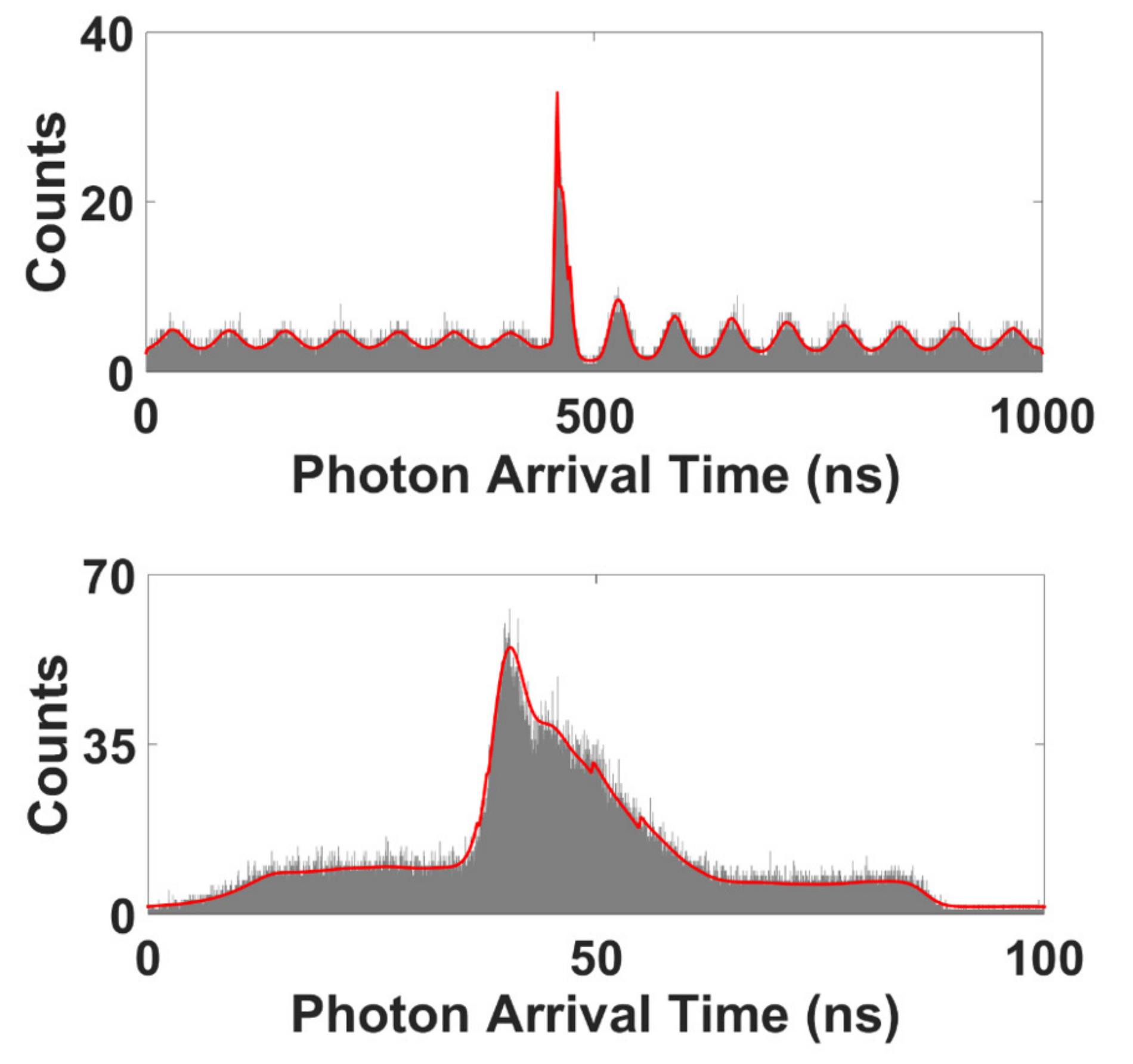}
\caption{Modified Bessel distribution (top) or arbitrary distribution (bottom). We note that our QRN generator can produce samples in real--time from any given distribution with high speed and accuracy. This figure is a reproduction of Figure 2 in \cite{nguyen2018programmable}.}
\label{perfQRN2}
\end{figure}

The NIST tests \cite{rukhin2001statistical} and Dieharder tests\footnote{Dieharder tests are presented here: \href{http://webhome.phy.duke.edu/~rgb/General/dieharder.php}{http://webhome.phy.duke.edu/~rgb/General/dieharder.php}}\footnote{Linux manual for Dieharder routines is available here: \href{https://linux.die.net/man/1/dieharder}{https://linux.die.net/man/1/dieharder}} were applied to our QRNs. In order to repeat a test that returns ``weak", we used the option \texttt{-Y 1} in the Dieharder routine, also known as ``resolve ambiguity mode". We observed that our QRNs exhibit the desired statistical properties. NIST and Dieharder tests were applied (see Figure \ref{perfQRN} for a summary of results). As it turns out, our QRNs pass all Dieharder tests for randomness, with all the p--values consistently above the significance level with high confidence. Moreover, our QRN generator can produce samples with any preassigned distribution as illustrated in figure \ref{perfQRN2}, where we show the histograms of QRNs in a modified Bessel distribution and an arbitrary distribution, respectively. 

We present in Table \ref{tab:comparisonANU} a comprehensive comparison between QuEST QRNs, ID Quantique (IDQ)\footnote{Swiss company ID Quantique official website: \href{http://www.idquantique.com/}{http://www.idquantique.com/}}, and Australian National University (ANU) Quantum Random Numbers Server\footnote{Australian National University Quantum Random Numbers Server: \href{https://qrng.anu.edu.au/}{https://qrng.anu.edu.au/}} QRNs. The random sample from IDQ and ANU were obtained via publicly available sources. In addition, several QRN samples were obtained by using the ID Quantique device Quantis-USB Model: USB-4M.V.13.11.08. Note that the QRNs generated at SIT outperform their counterparts, since our random numbers pass all the required statistical tests. By comparison, the IDQ numbers fail some Dieharder tests. The ANU numbers have similar quality as SIT QRNs, the only major difference being the generation speed (ANU has 5.7 Gbits/sec, while we can generate at 8 Gbits/sec). Moreover, our QRN generator can generate samples in any pre-specified distribution. In particular, generating normally distributed samples is straightforward from the experimental point of view. Furthermore, the functionality of our device can be extended to ensure secure encryption, fast communication, and more reliable and efficient Monte Carlo simulations.

\begin{table}
\centering
{\small
\begin{tabular}{lcccccc}
\hline
Source&	Data Size 	& Sample Size&	Range	&Distribution&	NIST&	Dieharder \\
         & (MB) &  (32-bit int)   &   &  & pass rate&pass rate\\
\hline
QuEST Lab	&830.9 &	2.077 $\x 10^8$	&[35, 4.295 $\x10^9$]&	Uniform	& 	15/15&	31/31\\
ANU 1&	1050	&2.621 $\x 10^8$&	[20, 4.295 $\x10^9$]	&Uniform		&15/15&	31/31\\
ANU 2&104.9 & 0.262$\x 10^8$ & [143, 4.295 $\x 10^9$] & Uniform  & 15/15& 28/31\\
IDQ (web data)&	125	&0.3125 $\x 10^8$&	[38, 4.295 $\x10^9$]	&Uniform&	15/15&	28/31\\
IDQ Quantis Device 1	&100&	0.25 $\x 10^8$&	[13, 4.295 $\x10^9$]	&Uniform&	15/15&	29/31\\
IDQ Quantis Device 2	&100&0.25 $\x 10^8$&  [118, 4.295 $\x 10^9$] & Uniform &	15/15&	28/31\\
\hline
\end{tabular}}
\caption{Comparison of our QRNs with their ANU and IDQ counterparts in terms of randomness tests.}
\label{tab:comparisonANU}
\end{table}

\section{Fractional Brownian motion}
We present in this section some well-known properties of the fractional Brownian motion (fBM), and we implement several simulation algorithms using both pseudo and quantum random numbers.

\subsection{Properties}
Let $W_t$ and $B_t,\ t\ge0$ be two independent Brownian motions (fBM), and denote with $W_t^H$ and $B_t^H$ the fBMs driven by $W_t$ and $B_t,$ respectively. By definition, a fractional Brownian motion (fBM) with Hurst parameter $H\in(0,1)$ is a zero mean Gaussian process $W^H=\{W_t^H,t\ge0\}$ with a temporal covariance function given by
\bel{covfBM}
\gamma(s,t)=\mathbb{E}\left[W^H_sW^H_t\right]=
\frac{1}{2}\left( t^{2H}+s^{2H} - |t-s|^{2H}\right],\ \forall s,t\ge0.
\ee
This process was first introduced by Kolmogorov in \cite{kolmogorov1940wienersche} and later studied by Mandelbrot and Van Ness in the pioneering work \cite{mandelbrot1968fractional}. The fBM is defined via the following stochastic representation:
\bel{srfBM}
W^H_t:=\frac{1}{\Gamma\left(H+\frac{1}{2}\right)}
\left[
\int_{-\infty}^{0}\left(    (t-s)^{H-1/2} -(-s)^{H-1/2} \right) \d W_s + 
\int_{0}^{t}    (t-s)^{H-1/2}\d W_s,
\right]
\ee
where $\{W_s,s\ge0\}$ is a Brownian motion and $\Gamma(\cdot)$ represents the Gamma function. Note that $W$ is recovered by taking $H=\frac{1}{2}$ in \eqref{srfBM}. We further remark that this
representation in terms of an integral with respect to Brownian motion is non-unique; see e.g. \cite{samoradnitsky2017stable}. A representation in terms of the Molchan-Golosov kernel will also be used in the sequel. That is, 
\[
B_t^H = \int_0^t K(t,s) dB_s,
\]
where $K$ is the Molchan-Golosov kernel
\begin{equation}
K(t,s) = c_H (t-s)^{H-\frac{1}{2}}F\left(H-\frac{1}{2},\frac{1}{2}-H,H+\frac{1}{2};1-\frac{t}{s}\right)\mathbf{1}_{[0,t]}(s), \label{eqn:molchon-golosov}
\end{equation}
with $$c_H=\left[\frac{2H\Gamma\left(\frac{3}{2}-H\right)}{\Gamma(2-2H)\Gamma\left(H+\frac12\right)}\right]^{1/2},$$ and $F$ is the Gauss hypergeometric function.

It can be shown that a normalized fractional Brownian motion $\{W^H_t\}_{t\ge0}$ with $H\in(0,1)$ is uniquely characterized by the following properties:
\begin{enumerate}[(i)]
\item $W^H_t$ has stationary increments
\item $W^H_0 = 0,\quad
\E[W^H_t] = 0,\quad 
\text{Var}(W^H_t) = \E[(W^H_t)^2] = t^{2H},\ \forall t\ge0
$ 
\item $W^H_t$ has a Gaussian distribution for $t>0$.
\end{enumerate}
From \eqref{covfBM} we deduce that $$\E[{|W_t^H-W_s^H|}] = |t-s|^{2H}$$ and, as a consequence, the trajectories of $W^H$ are almost surely locally $\alpha-$H\"older continuous for all $\alpha\in(0,H)$. Since $W^H$ is not a semi-martingale nor Markovian if $H\ne1/2$, we cannot use the stochastic calculus of It\^o with respect to the fBM. Over the past years, some new techniques have been developed in order to
define stochastic integrals with respect to a fBM, e.g. the Skorokhod integral defined in the framework of the
Malliavin calculus \cite{nualart2006malliavin}, \cite{alos2001stochastic}, or the pathwise Riemann-Stieltjes integral \cite{zahle1998integration}.

\subsection{Simulation and comparison}
A plethora of methods are available for simulating sample paths of fBMs, and we refer the reader to \cite{dieker2004simulation} for a good review of properties and algorithms. Let $$\Pi=\{0=t_0,t_1,\dots,t_n=T\}$$such that $t_0<t_1<\cdots<t_n$ be a partition of the fixed time interval $[0,T]$. We consider here two methods for testing purposes:
\begin{enumerate}[(i)]
\item The Davies and Harte method for generating fBM samples $\{W^H_{t_0}, W^H_{t_1},\dots,W^H_{t_n}\}$ on the partition $\Pi$ was implemented. The algorithm was originally proposed by \cite{davies1987tests} and it is presented in detail in Section 2.1.3 of \cite{dieker2004simulation}. 

\item The hybrid scheme for Brownian semistationary processes given in \cite{bennedsen2017hybrid} was used to generate sample paths for fBMs $\{ B^H_{t_k},\ k=0,1,2,\dots,n\}$ using the Molchan--Golosov kernel. The algorithm is based on discretizing the stochastic integral representation of the process in the time domain. More recently, \cite{mccrickerd2017turbocharging} considered methods for turbocharging Monte Carlo pricing under the so--called rough Bergomi model. 
\end{enumerate}

\begin{figure}
\centering
\includegraphics[scale = 0.43]{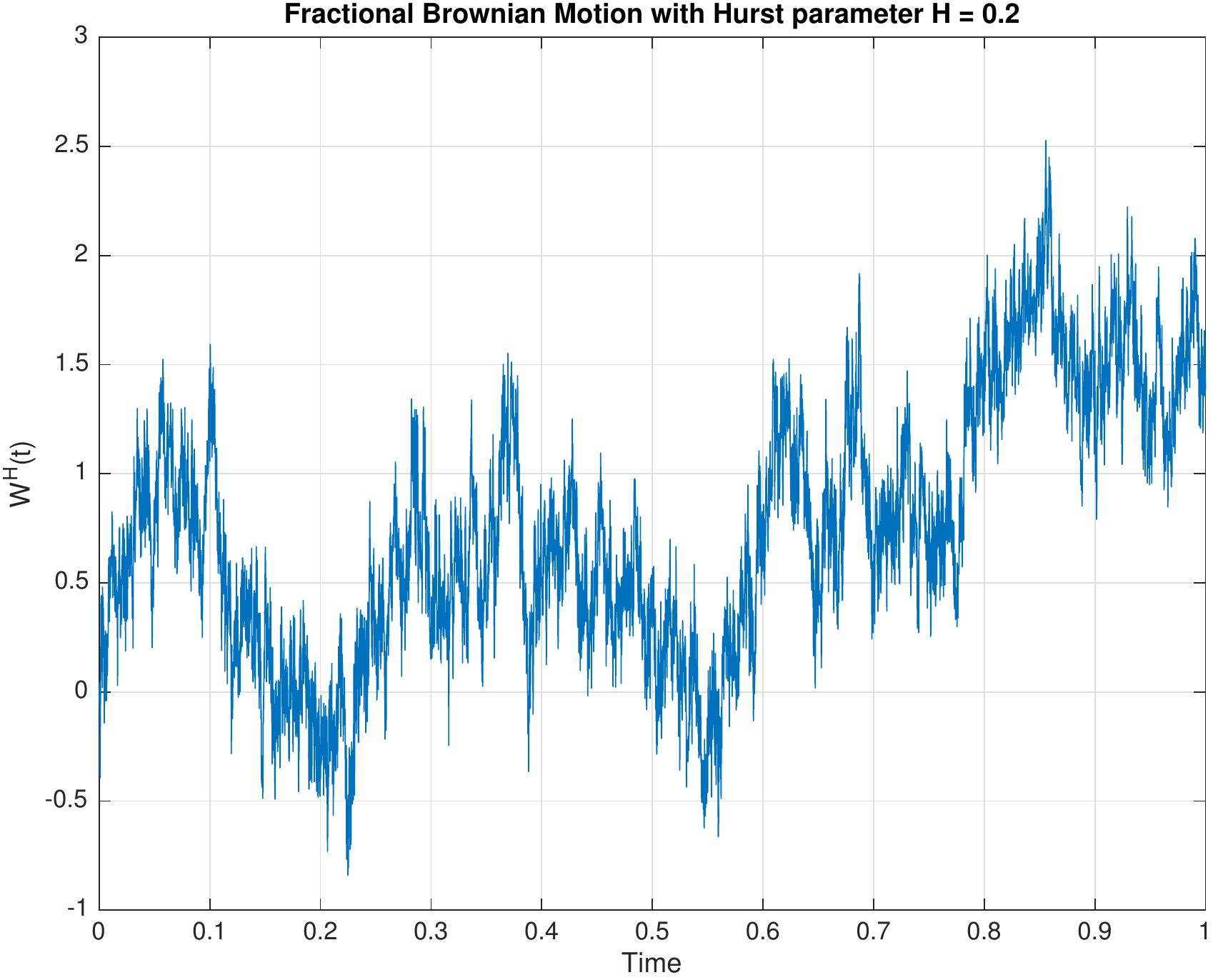}
\includegraphics[scale = 0.43]{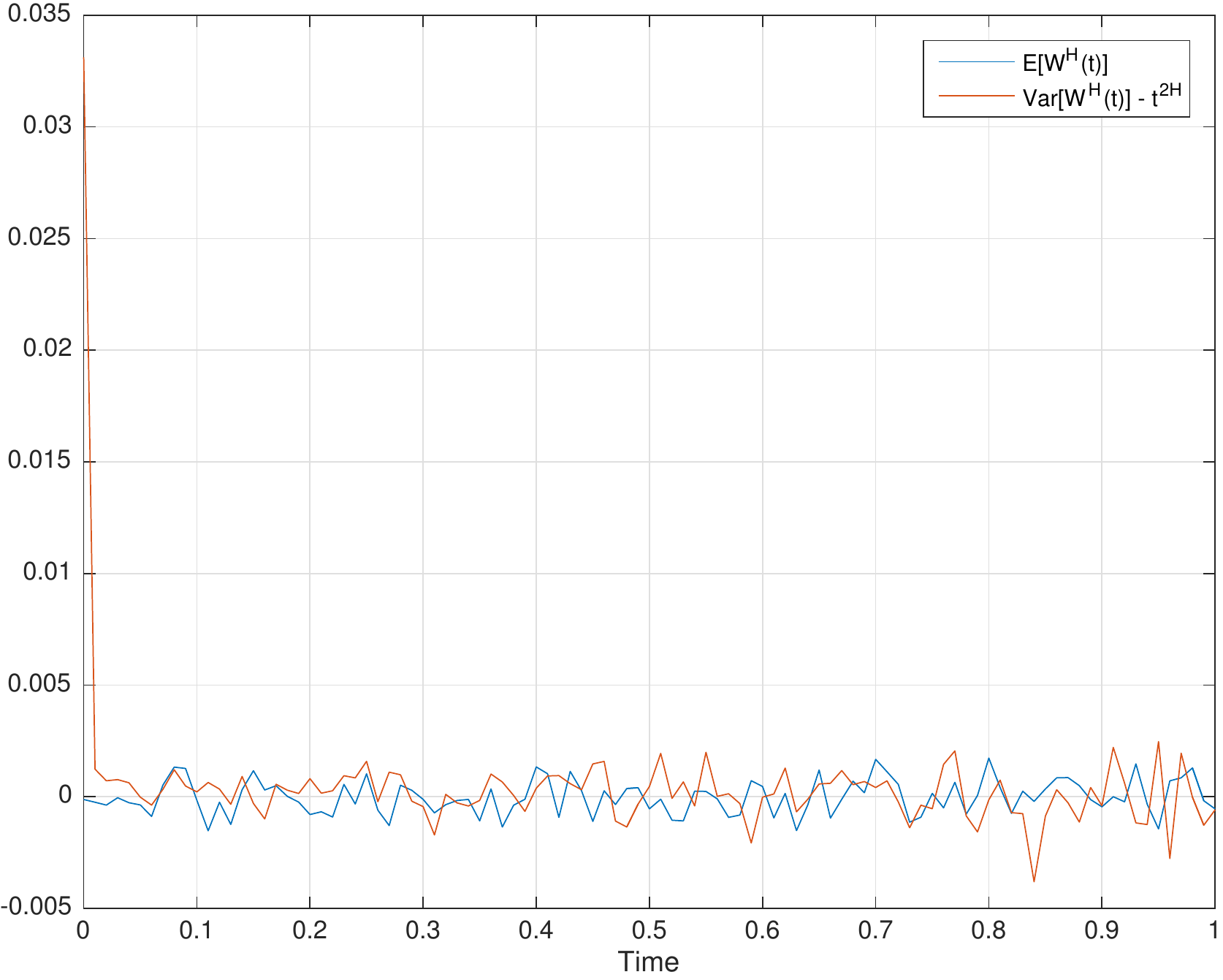}
\caption{{One path of simulated fBM via Davies and Harte method (left) and its variance structure using PRNs (right). Parameters are $H = 0.2, T = 1$ year, $n=252$ samples, $N=10^6$ paths.}}
\label{figfBM}

\
\\

\includegraphics[scale = 0.51]{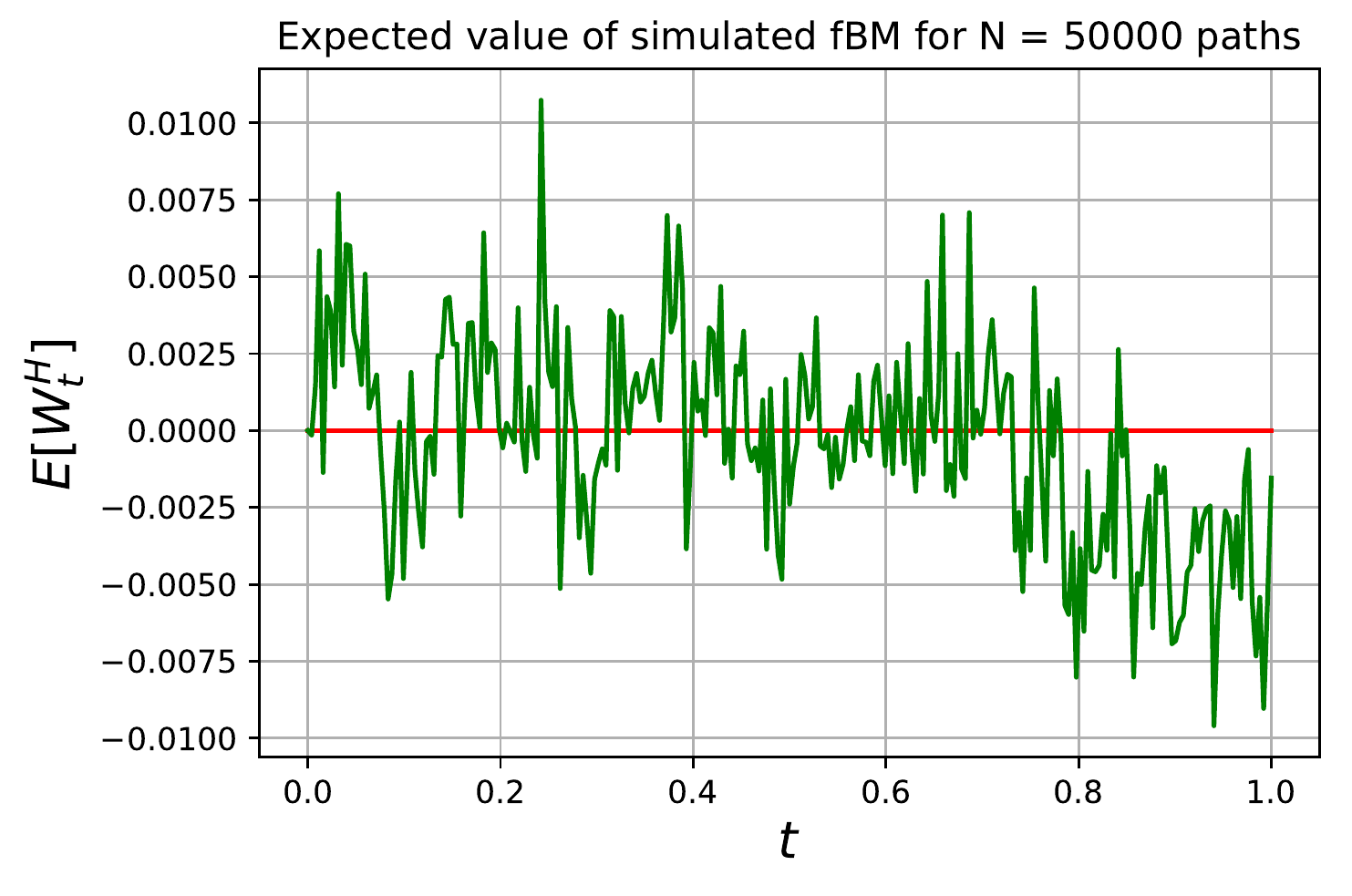}
\includegraphics[scale = 0.51]{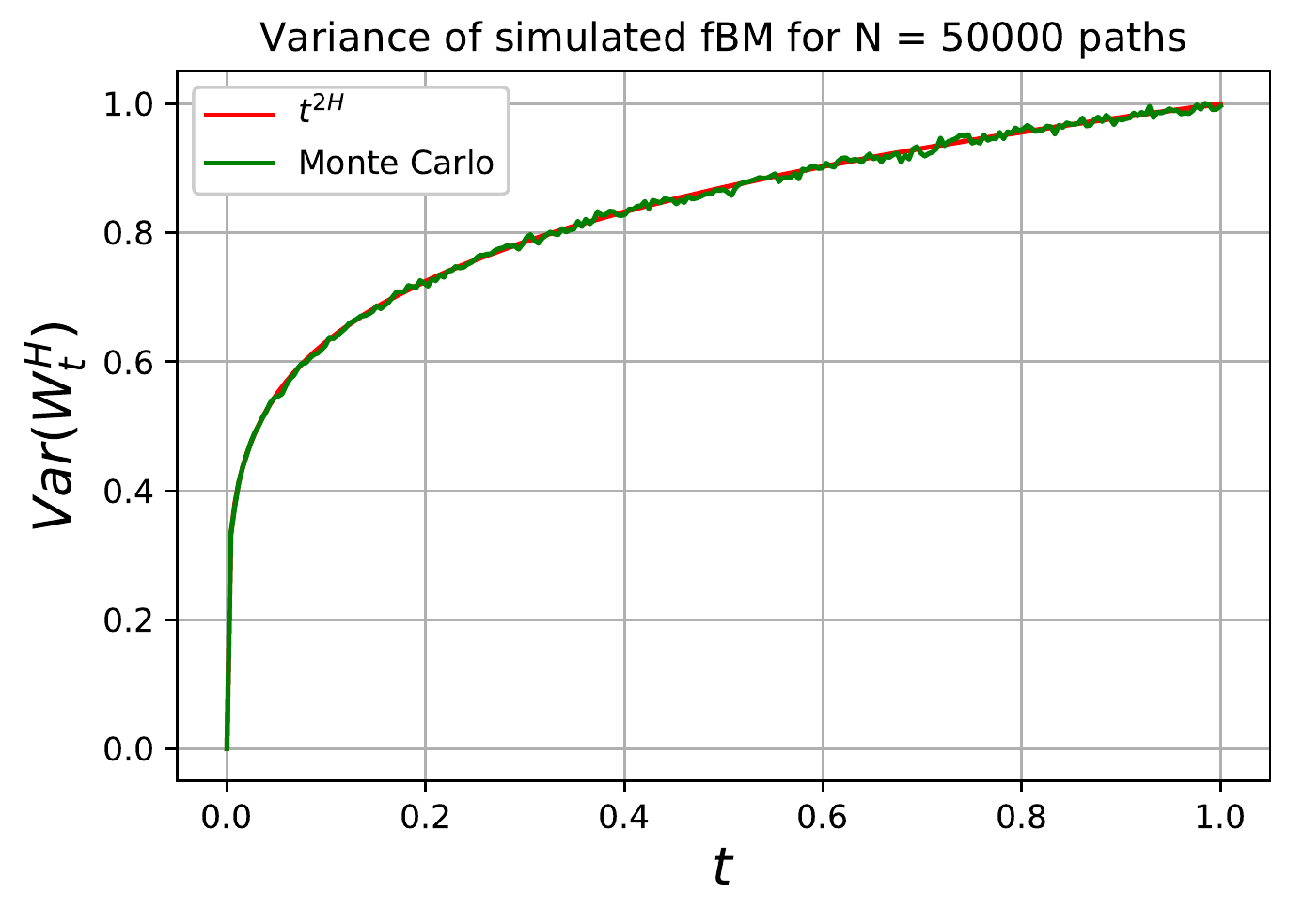}
\caption{Mean (left) and variance (right) for fBM paths generated via the hybrid scheme and PRNs. Parameters are $H = 0.2, T = 1$ year, $n=252$ samples, $N=50,000$ paths.}
\label{fig:evar}

\
\\

\includegraphics[scale = 0.59]{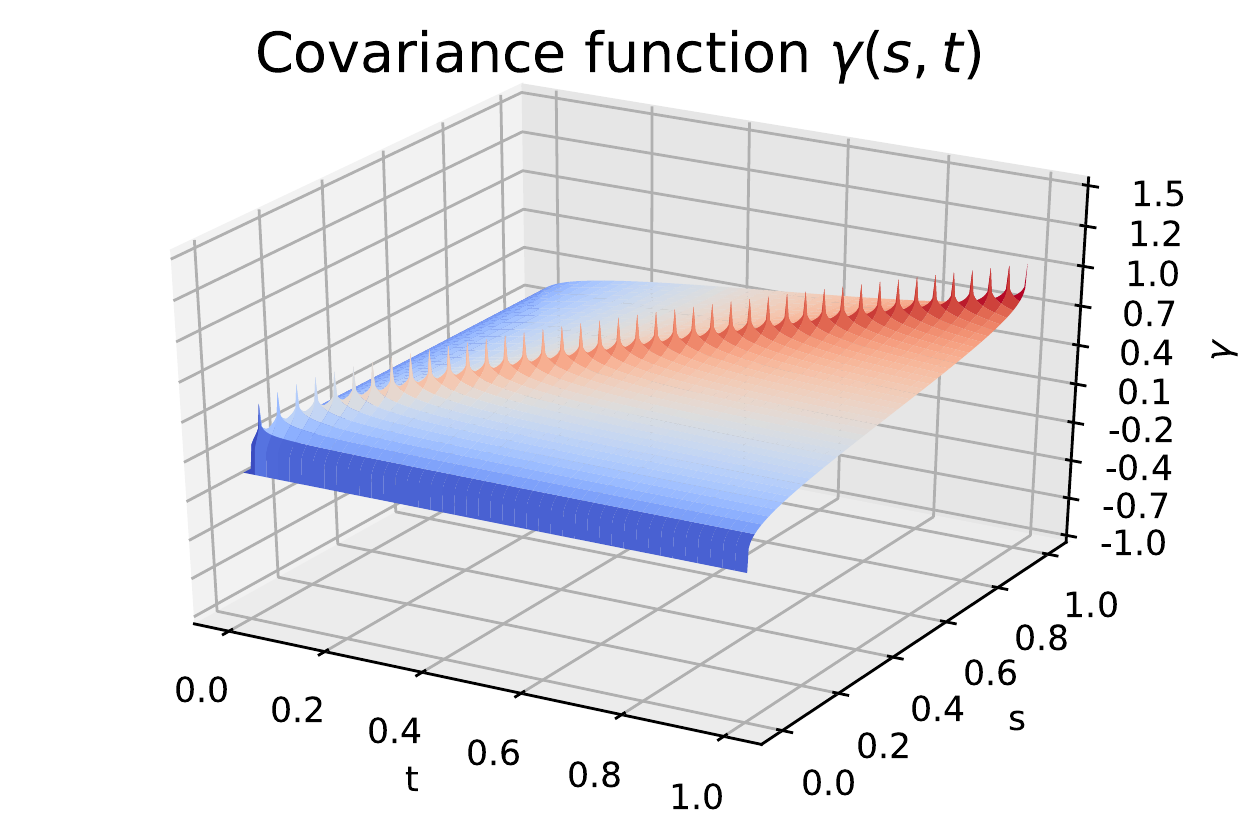}
\includegraphics[scale = 0.59]{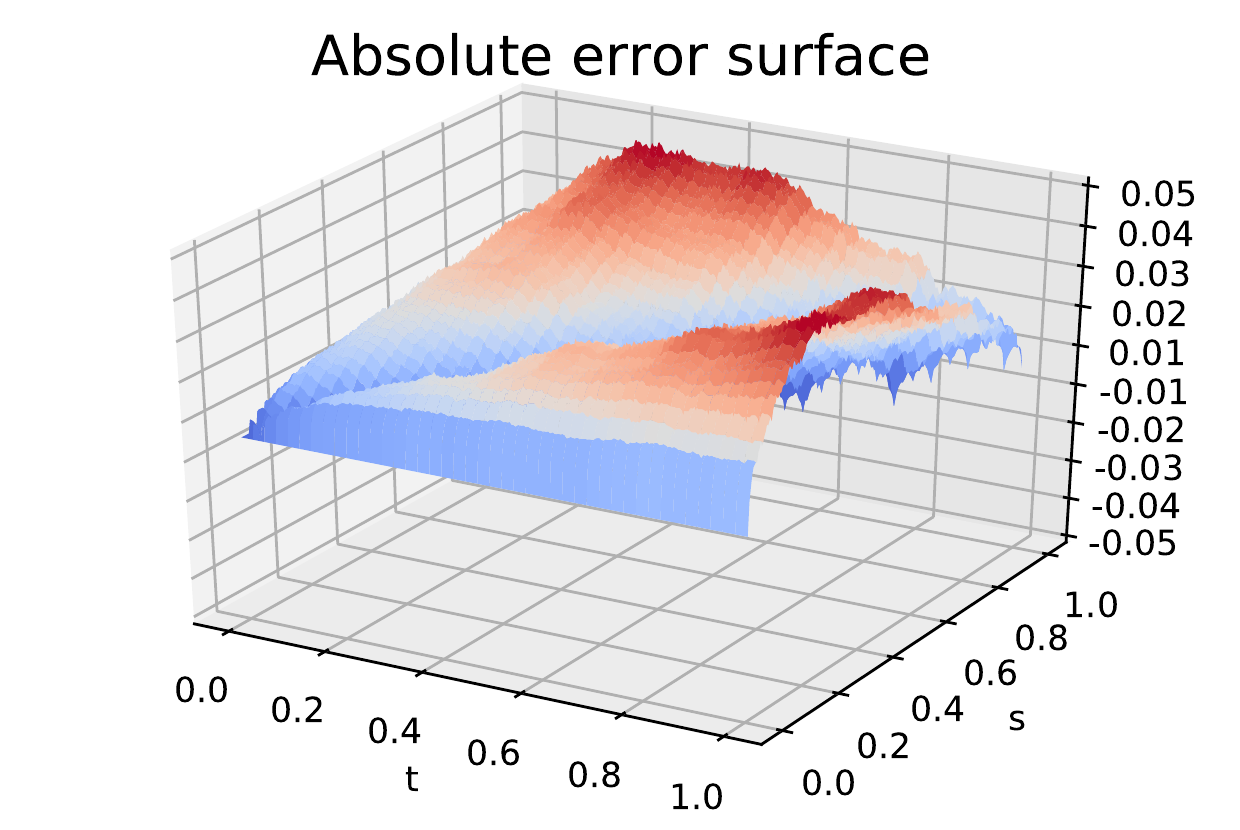}
\caption{Covariance function of fBM $\gamma(s,t):=\frac{1}{2}\left(t^{2H}+s^{2H} - |t-s|^{2H}\right)$ (left) and error $\gamma(s,t) - \gamma^{MC}(s,t)$ (right) on the domain $\Pi\times\Pi$. Parameters are $H = 0.2, T = 1$ year, $n=252$ samples, $N=50,000$ paths.}
\label{fig:covFBM}
\end{figure}

In order to evaluate the algorithms, several test routines calculating moments and errors for fractional processes have been implemented: mean $\mu$ and variance $v$ as a function of time via Monte Carlo simulations, a chi--square test for fractional Gaussian noise \cite{dieker2004simulation}, \cite{beran1994statistics}, as well as the 2D correlation structure $\gamma^{MC}$ via sample paths: 
\begin{eqnarray}
\mu_k\equiv\mu(t_k) &:=& \frac{1}{N} \sum_{i=1}^N B^H_i (t_k),\\
v_k\equiv v(t_k) &:=& \frac{1}{N-1} \sum_{i=1}^N \Big(B^H_i (t_k) - \mu(t_k)\Big)^2,\\
\gamma^{MC}_{kj}\equiv\gamma^{MC}(t_k,t_j) &:=& \frac{1}{N-1} \sum_{i=1}^N \Big(B^H_i (t_k) - \mu(t_k)\Big)\Big(B^H_i (t_j) - \mu(t_j)\Big)
\end{eqnarray}
where $k,j = 0,1,2,\dots,n$ and $N$ is the number of MC paths. The results are presented in Figure \ref{figfBM} (Davies and Harte method), and Figures \ref{fig:evar} and \ref{fig:covFBM} (hybrid scheme). By inspection, we notice that the generated sample paths have the required properties that are specific to fBMs. 

\begin{figure}
\centering
\includegraphics[scale = 0.52]{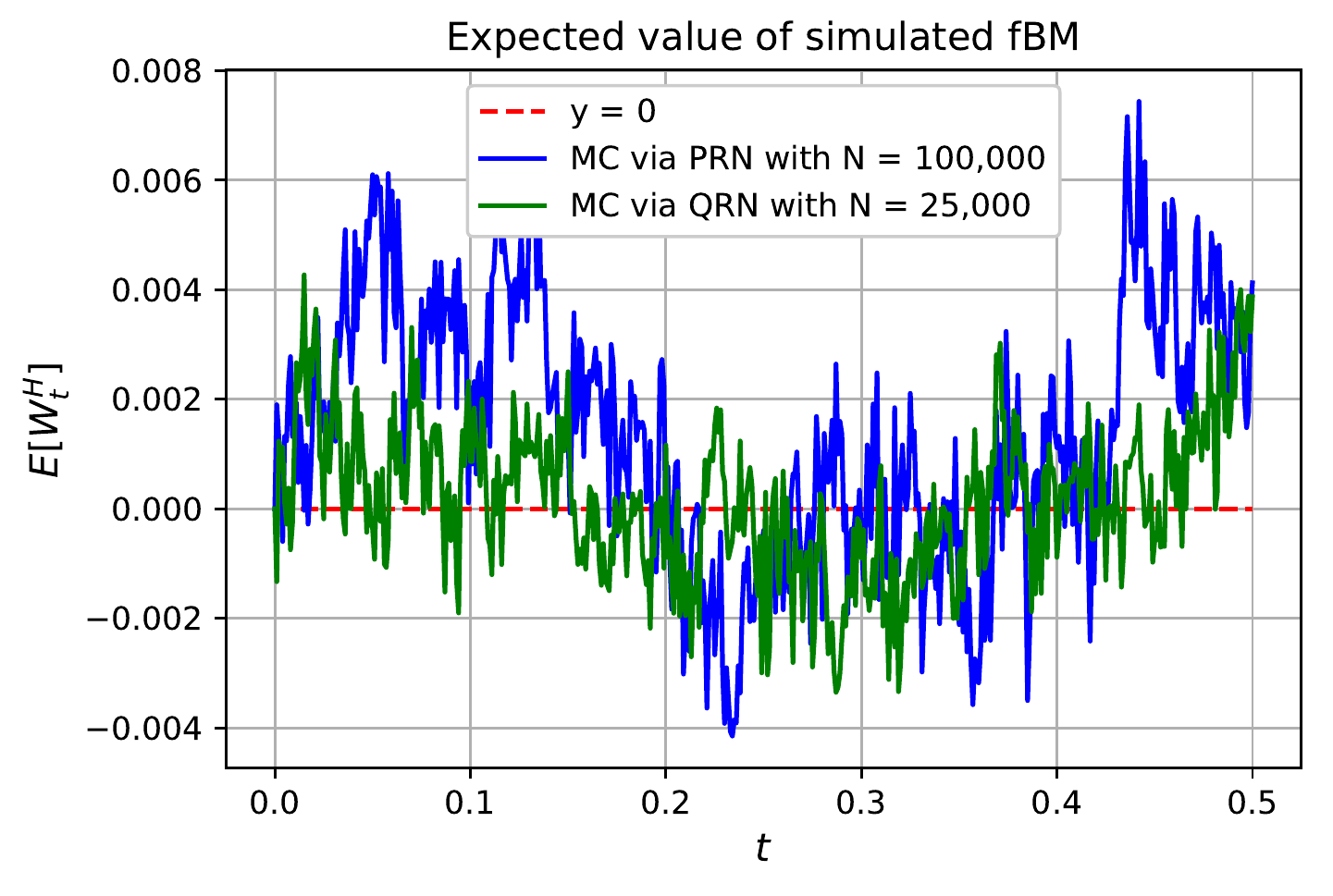}
\includegraphics[scale = 0.52]{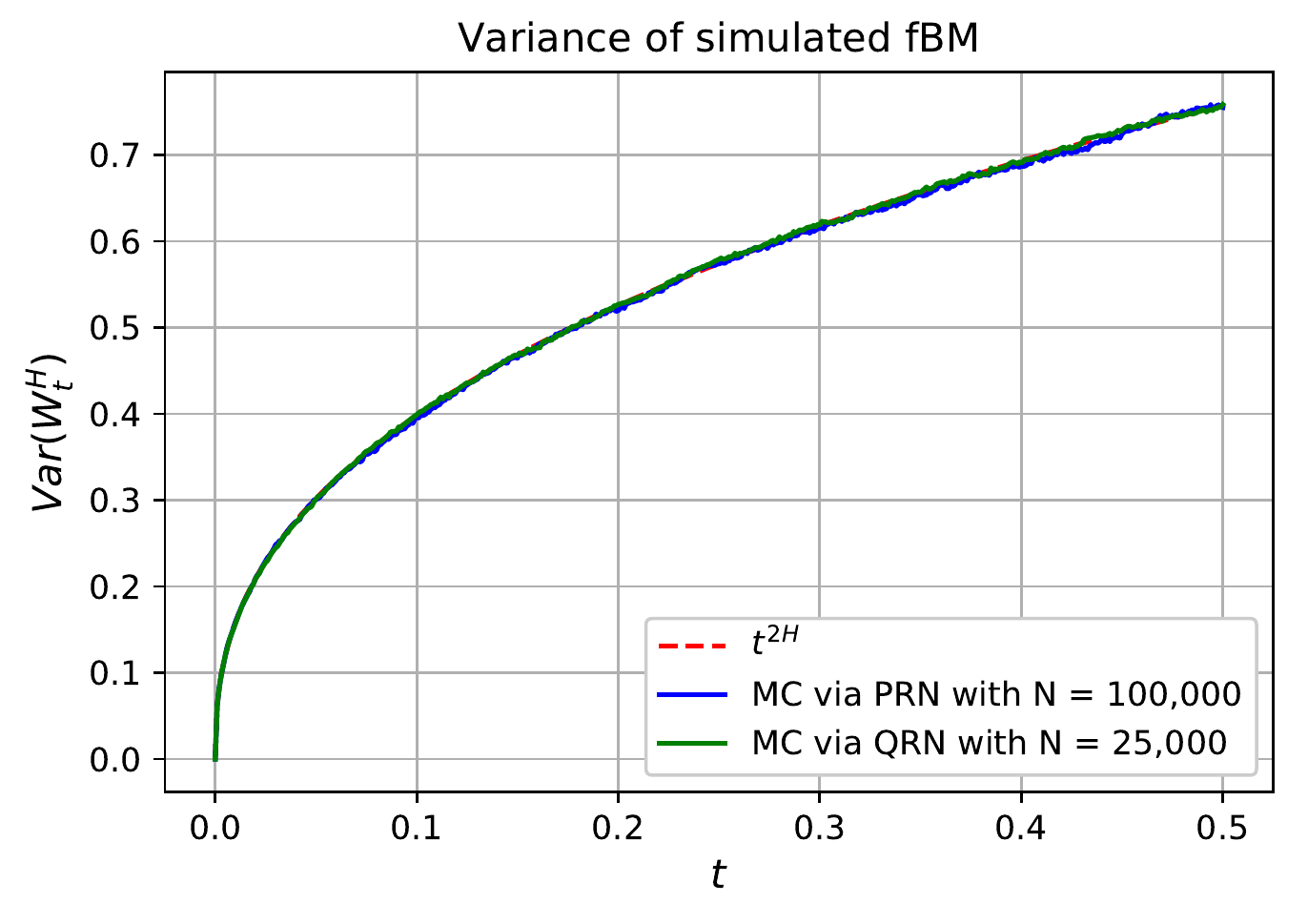}

\
\\

\includegraphics[scale = 0.52]{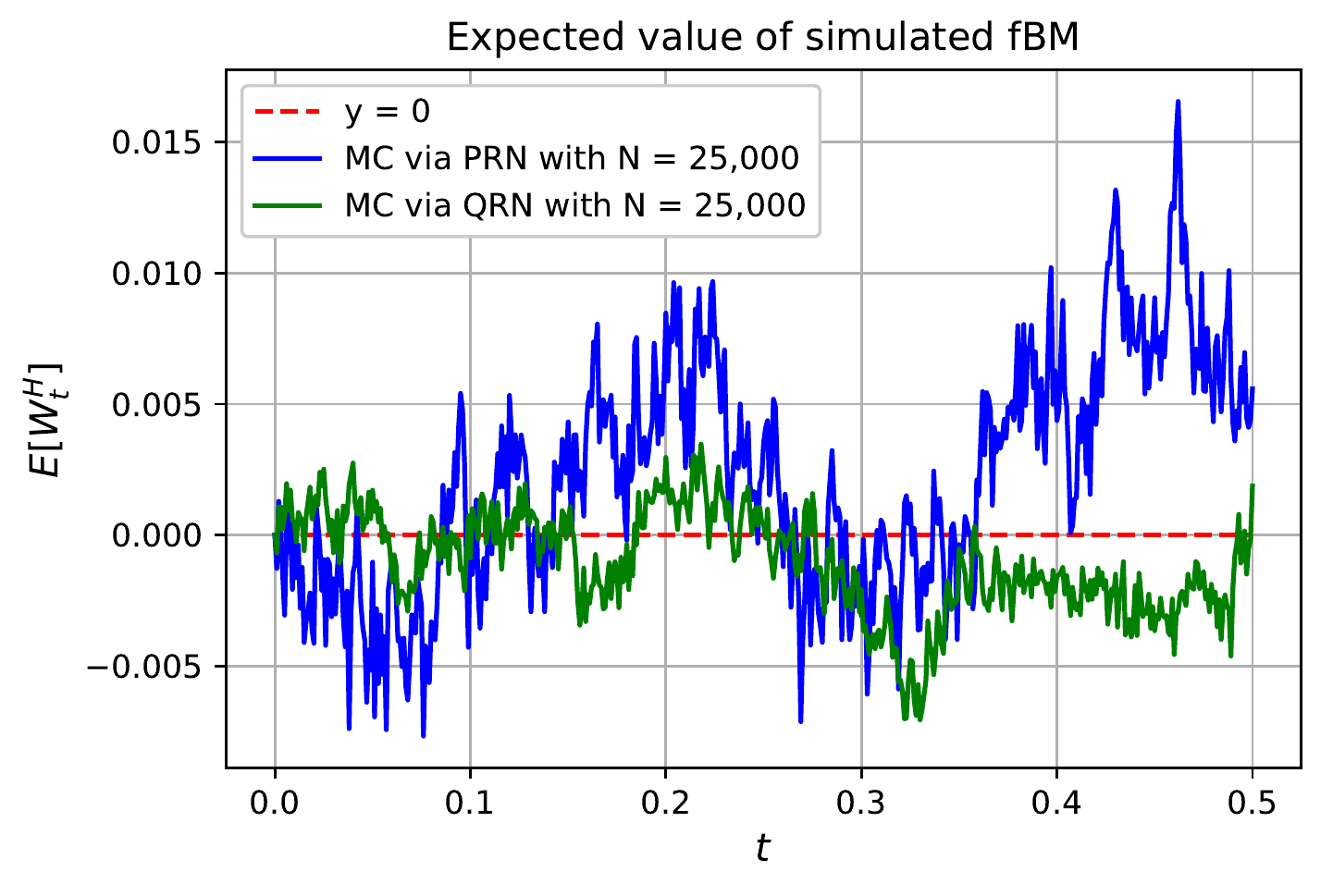}
\includegraphics[scale = 0.52]{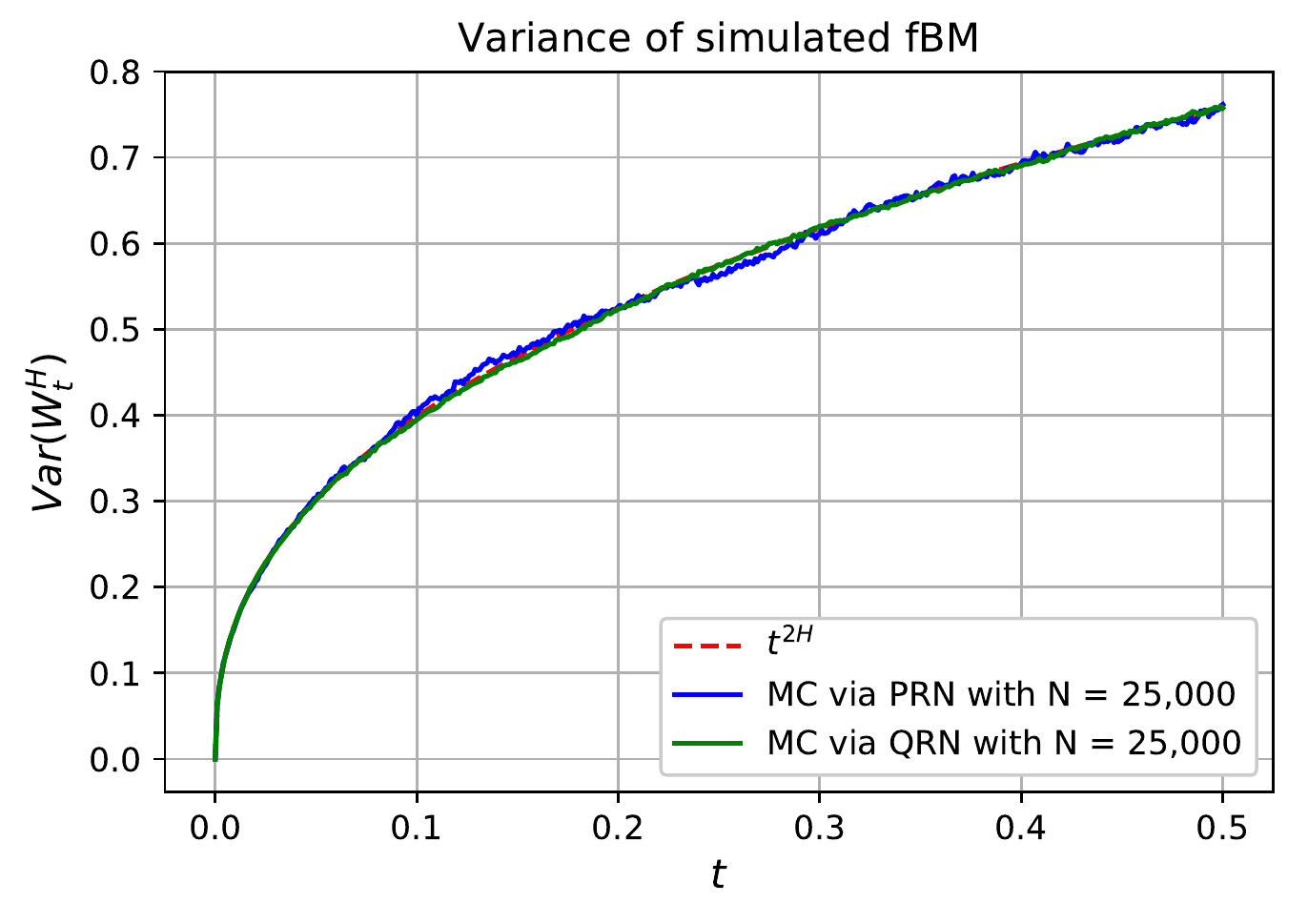}
\caption{Comparison between fBM statistical properties via PRNs and QRNs. It is easy to notice that similar accuracy may be obtained with a smaller number of MC trials.}
\label{fBMQRN}

\
\\

\includegraphics[scale=0.52]{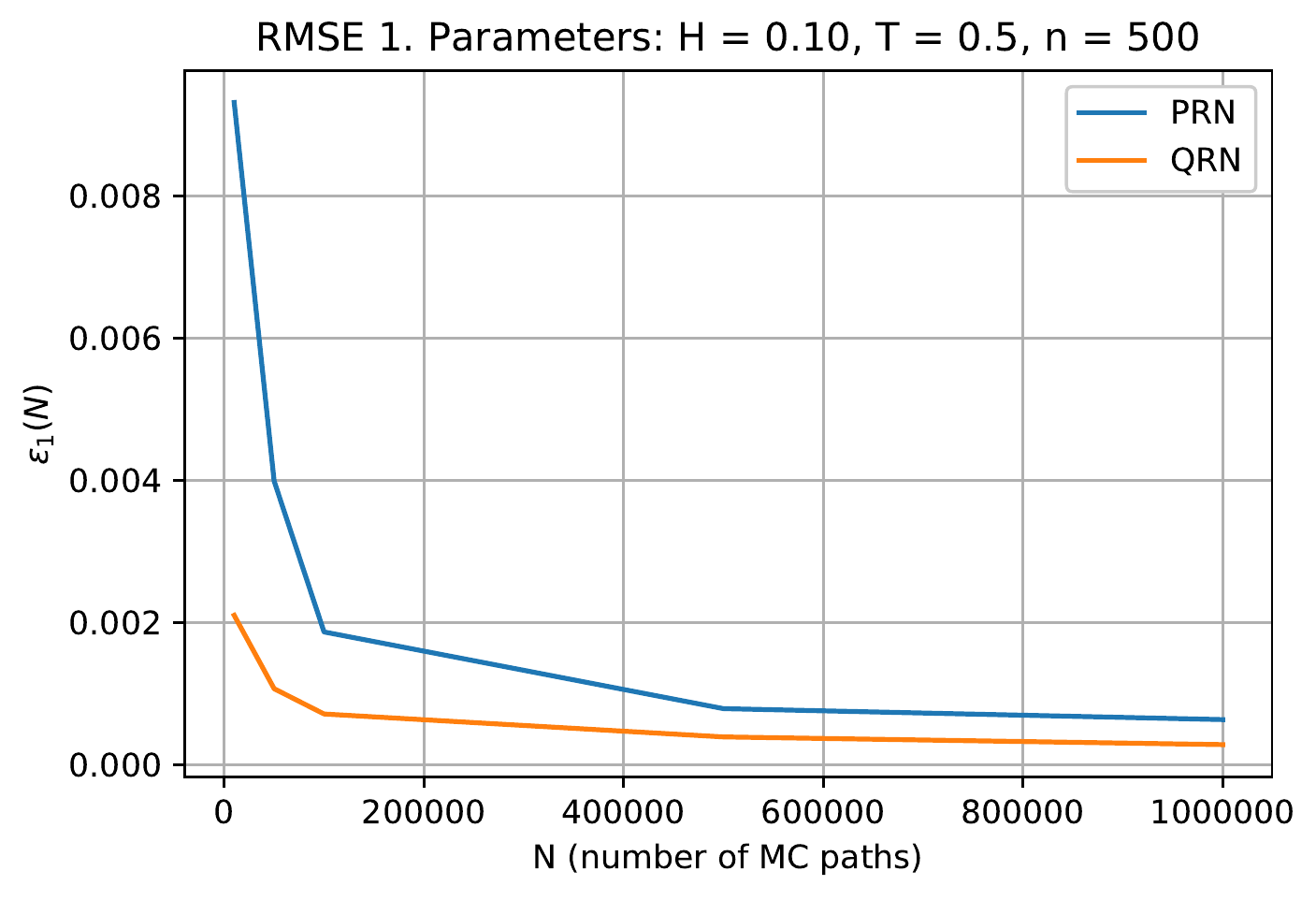}
\includegraphics[scale=0.52]{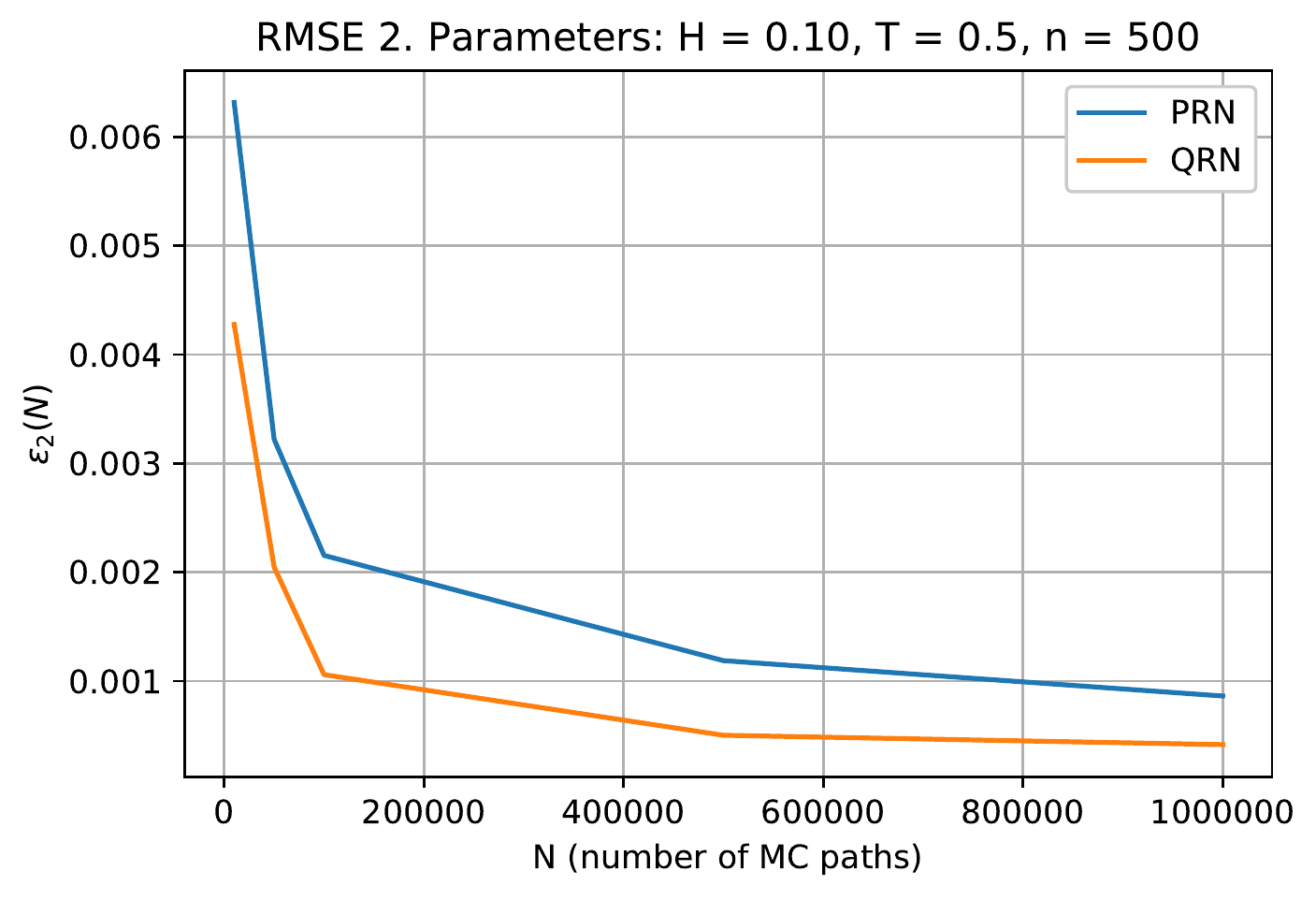}
\caption{RMSE for mean and variance on the partition $\Pi$ versus $N$. Plot corresponds to Table \ref{tab:errors}. Parameters are $H=0.1, T=0.5, n=500$.}
\label{fig:errors}
\end{figure}

We employ now for comparison purposes the same simulation methods using this time QuEST QRNs. Fractional BM paths are generated via the hybrid scheme using both QRNs and PRNs. Statistical properties for sample fBM paths using both PRNs and QRNs are shown in Figure \ref{fBMQRN}. To measure our QRNs performance, we introduce the following errors:
\begin{eqnarray}
\varepsilon_1 &=& \sqrt{\frac{1}{n+1}\sum_{k=0}^n \mu_k^2}\\\cr
\varepsilon_2 &=& \sqrt{\frac{1}{n+1}\sum_{k=0}^n \left(v_k - t_k^{2H}\right)^2}\\\cr
\varepsilon_3 &=& \sqrt{\frac{1}{(n+1)^2}\sum_{k=0}^n\sum_{j=0}^n \left( \gamma^{MC}_{kj} - \gamma(t_k,t_j)\right)^2}
\end{eqnarray}

Note that $\varepsilon_{i},i=1,2,3$ are the root mean squared errors with respect to the first two moments and covariance in time of our simulated fBM paths. We present in Table \ref{tab:errors} the errors for different numbers of MC paths $N$ and different source of randomness (quantum versus classical). The corresponding plots are given in Figure \ref{fig:errors}. We note that the QRN rate of convergence is superior to its classical counterpart, thus yielding similar performance with less random samples. Moreover, it is clear that the execution time is smaller when using QRNs, since no post-processing algorithm is applied (as opposed to PRNs, where Box-Muller or Ziggurat transformations need to be applied) and the random numbers are fed in real--time to the path generation algorithm. The function $\varepsilon_3(N)$ is almost constant, but we note once again that QRNs provide superior results.

\begin{table}
\centering
\begin{tabular}{clccc}
\hline
Source of randomness& $N$& $\varepsilon_1$ & $\varepsilon_2$ & $\varepsilon_3$ \\  
\hline
& 10,000 & 0.00931595& 0.00631624 & 0.08907148\\
& 50,000 & 0.00399428 & 0.00322211 & 0.08883075\\
PRN & 100,000 & 0.00187179 & 0.00215443 & 0.08904862\\
& 500,000 & 0.00079351 & 0.00118751& 0.08925236\\
& 1,000,000 &  0.00063936 & 0.00063936 & 0.08935881\\
\hline
& 10,000 & 0.00210541& 0.00427726 & 0.030099337\\
& 50,000 & 0.00107404 & 0.00204733 & 0.030027625\\
QRN & 100,000 & 0.00071814 & 0.00105875 & 0.030092520\\
& 500,000 & 0.00039584 & 0.00050180& 0.030153221\\
& 1,000,000 &  0.00028751 & 0.00041583 & 0.030184910\\
\hline
\end{tabular}
\caption{Comparison between RMSEs for PRN versus QRN as a function of the number of MC paths $N$.}
\label{tab:errors}
\end{table}




\section{Applications to fractional stochastic processes}

We discuss now some interesting applications of our QRNs in financial engineering: the rough fractional stochastic volatility model and pricing of a volatility dependent option assuming the fractional SABR model.

\subsection{The Rough Fractional Stochastic Volatility model}

The analysis presented in this section is based on \cite{gatheral2018volatility}, where the authors propose the so--called Rough Fractional Stochastic Volatility (RFSV) model, which was shown to be compatible with the empirical smoothness of volatility in financial markets. Moreover, it was shown via empirical analysis that the log--volatility process estimated from high--frequency financial data behaves like a fBM with Hurst exponent of order 0.1.

Consider the following model from \cite{gatheral2018volatility}:
\bel{rfsvm}
\baa{rcl}
\sigma_t &=&\sigma_0 \exp\{X_t\}\\
\d X_t &=& \nu\d W^H_t - \alpha(X_t - m)\d t,\ X_0 = x_0>0,
\eaa
\ee
with $m\in\R$, and $\alpha,\nu,\sigma_0$ as positive parameters. Note that by choosing the fractional Ornstein-Uhlenbeck process (fOU) for the log volatility above ensures a stationary model. Recall that the fBM can be alternatively characterized by: 
\bel{fBMforfSV}
{
\left|W_{t+\Delta}^H-W_t^H\right|^q
}= K_q\Delta^{qH},
\ee
for any $t\in\R,\Delta\ge0,q>0$, where $K_q$ is the $q^{th}$-moment of the absolute value of a standard Gaussian variable.
\subsubsection{Full circle calibration and estimation}
\begin{figure}
\centering 
\includegraphics[scale = 0.42]{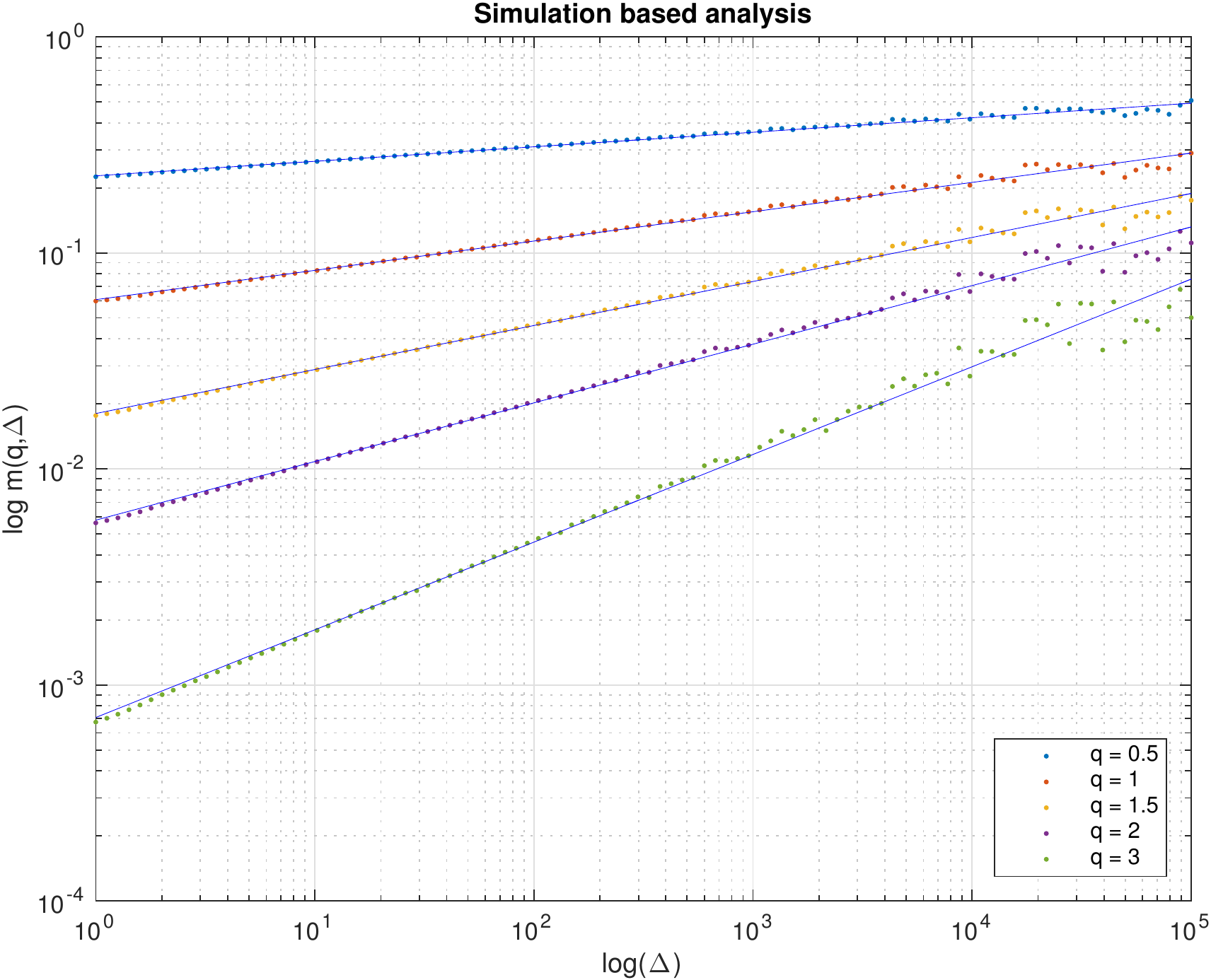}
\includegraphics[scale = 0.42]{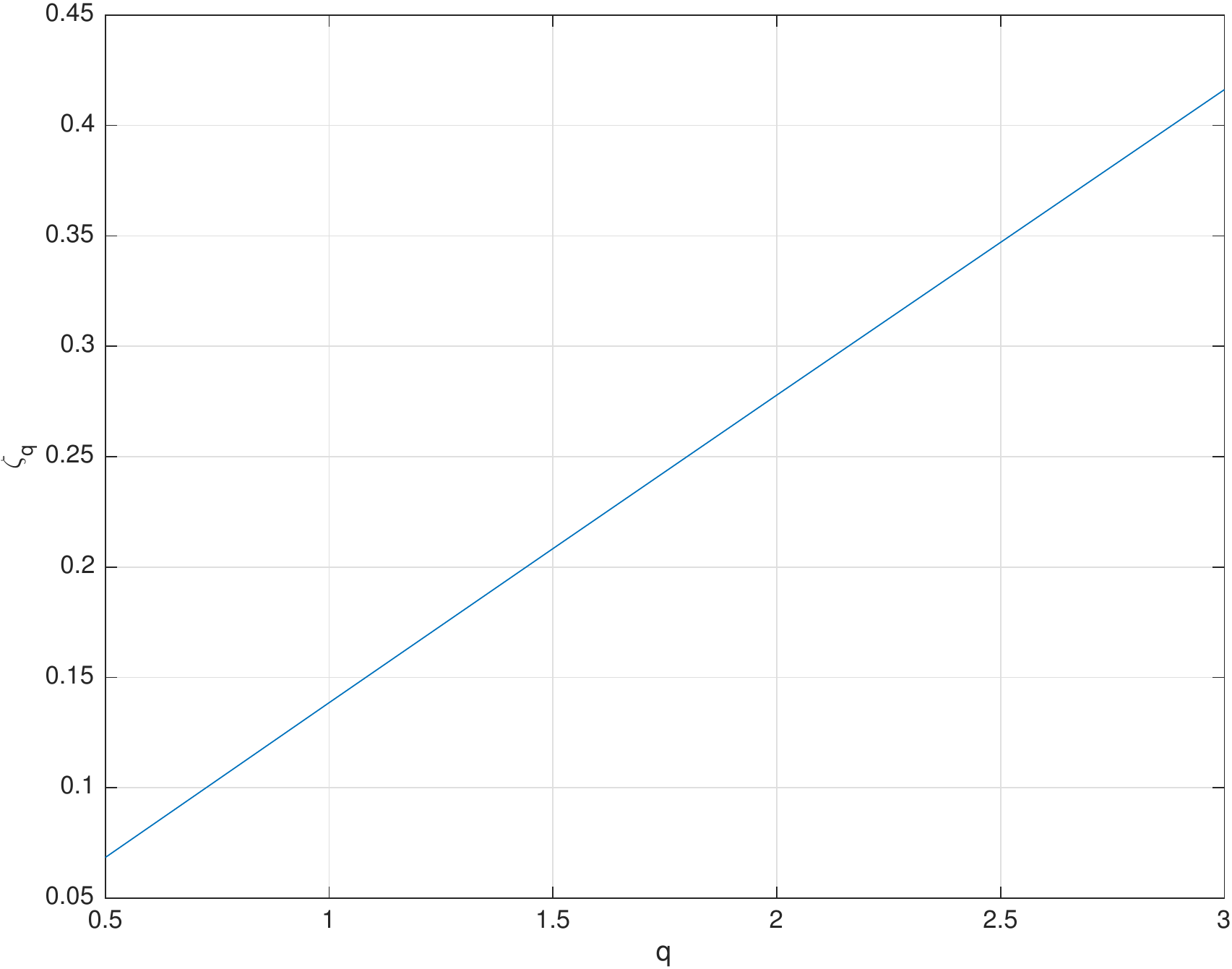}
\caption{\footnotesize{$\log m(q,\Delta)$ as a function of $\log\Delta$ for simulated data and estimating the Hurst parameter by using $\zeta_q\sim Hq$. We obtain H = 0.1354 (the true value is $H=0.14$)}}
\label{figLogLog}

\
\\

\
\\

\includegraphics[scale = 0.42]{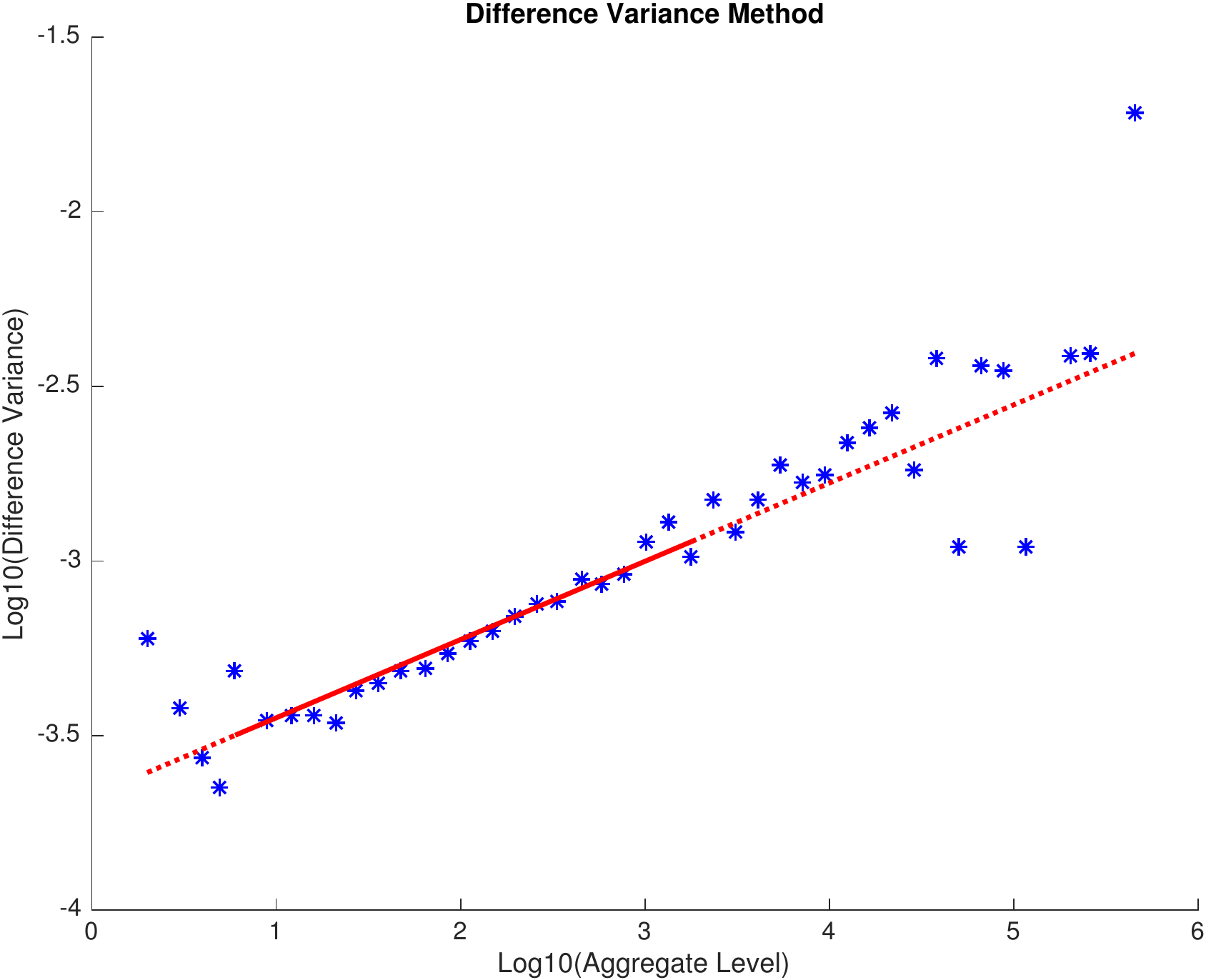}
\includegraphics[scale = 0.42]{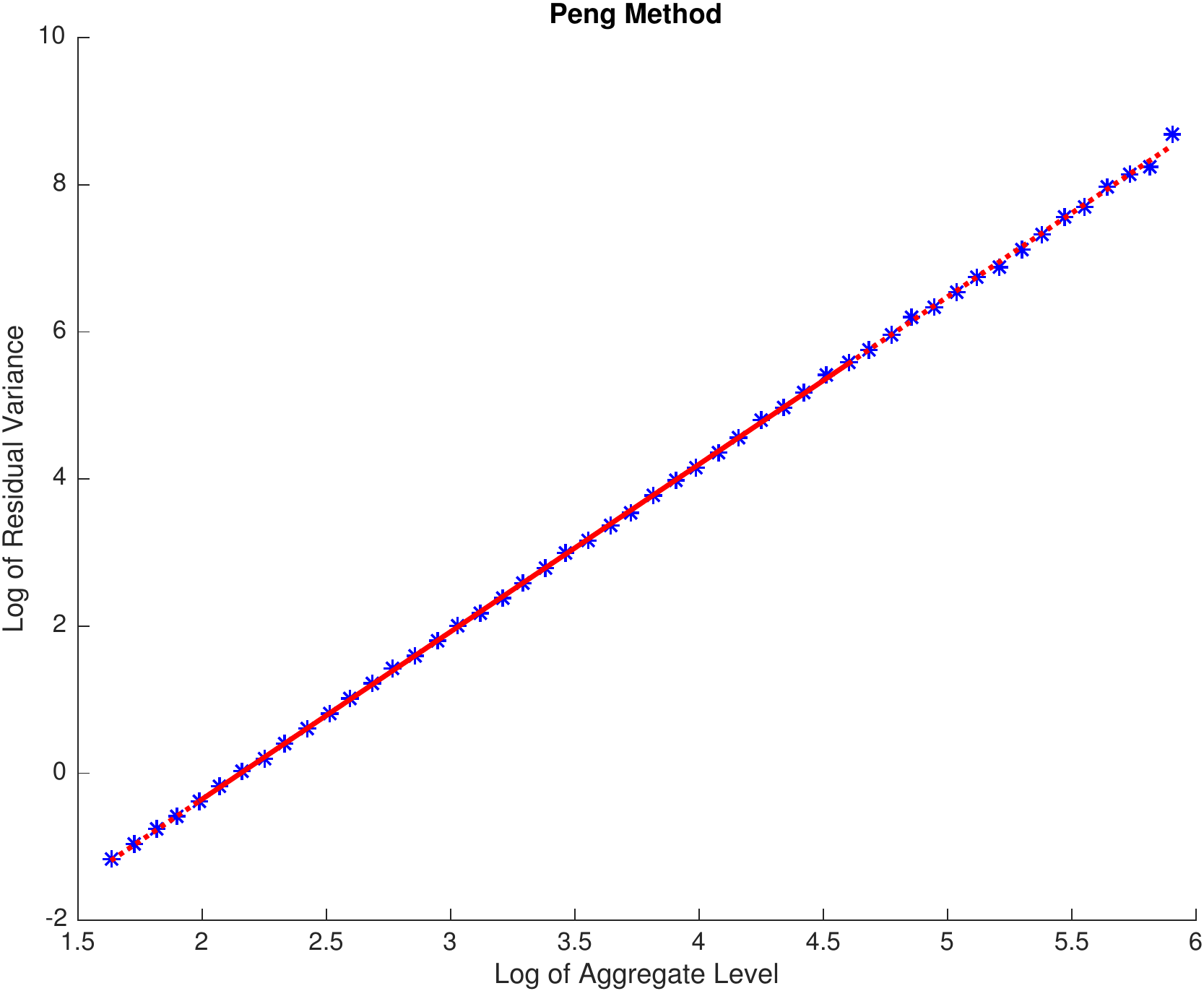}
\caption{\footnotesize{Estimating the Hurst parameter from $\log \sigma_t$ samples via difference variance method and Peng method. We obtain $H = 0.1121$ and $H = 0.1385$, respectively (the true value is $H=0.14$).}}
\label{figHE}
\end{figure}

We implement now a full circle calibration and estimation analysis of the RFSV model, which can be summarized in the following steps:
\begin{enumerate}[(i)]
\item Initialize the parameters and simulate one path of fBM using QuEST QRN generator.
\item Construct one path for a stationary fOU process $\{X_t\}_{t\ge0}$ and for the fractional stochastic volatility process $\{\sigma_t\}_{t\ge0}$ given in \eqref{rfsvm}.

\item Compute the function $m(q,\Delta)$ defined as:
\[
m(q,\Delta) = \frac{1}{N}\sum_{k=1}^N \left| \log(\sigma_{k\Delta}) - \log(\sigma_{(k-1)\Delta})\right|^q
\]
The plot of $\log m(q,\Delta)$ versus $\log\Delta$ for different values of $q$ is given in Figure \ref{figLogLog}. 

\item Note that for a given $q$, the points lie on a straight line. Thus the log volatility increments have the following scaling property in expectation:
\[
{\left| \log(\sigma_{k\Delta}) - \log(\sigma_{(k-1)\Delta})\right|^q} = K_q\Delta^{\zeta_q},\ k=1,2,\dots,N
\]
where $\zeta_q>0$ is the slope associated with $q$. Compare this with \eqref{fBMforfSV} and further note that $\zeta_q$ is linear with respect to $q$, i.e. $\zeta_q \sim qH$. Thus the slope of this line is the Hurst parameter. From our simulations, we obtained $H = 0.1354$ (in order to generate the sample path, we used $H=0.14$).  

\item We estimate the Hurst parameter from $\log \sigma_t$ samples via difference variance method and the Peng method, see \cite{dieker2004simulation}. We obtain $H = 0.1121$ and $H = 0.1385$, respectively. The results are plotted in Figure \ref{figHE}. 
\end{enumerate}


\subsubsection{Empirical Analysis}
\begin{figure}
\centering
\includegraphics[scale = 0.42]{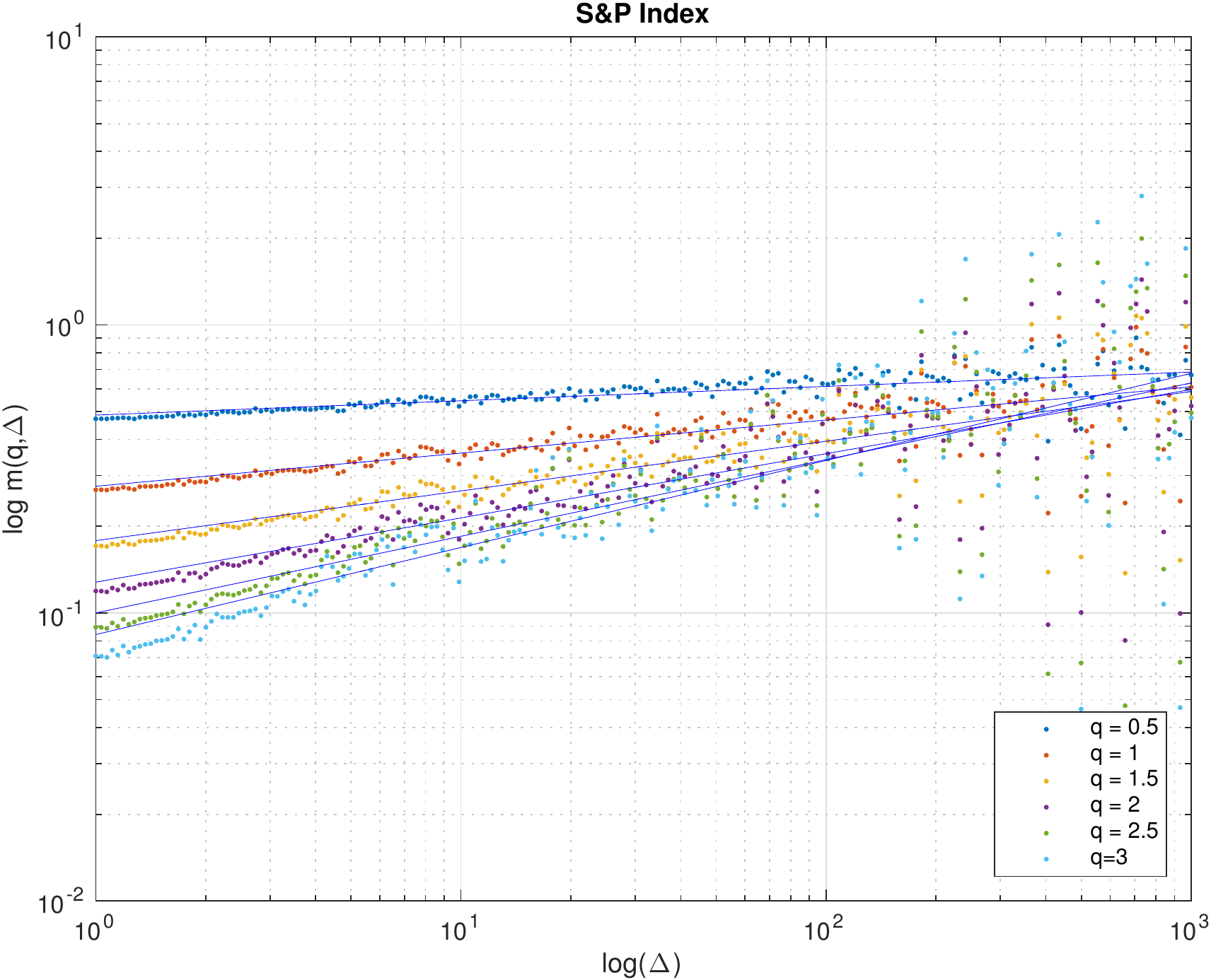}
\includegraphics[scale = 0.42]{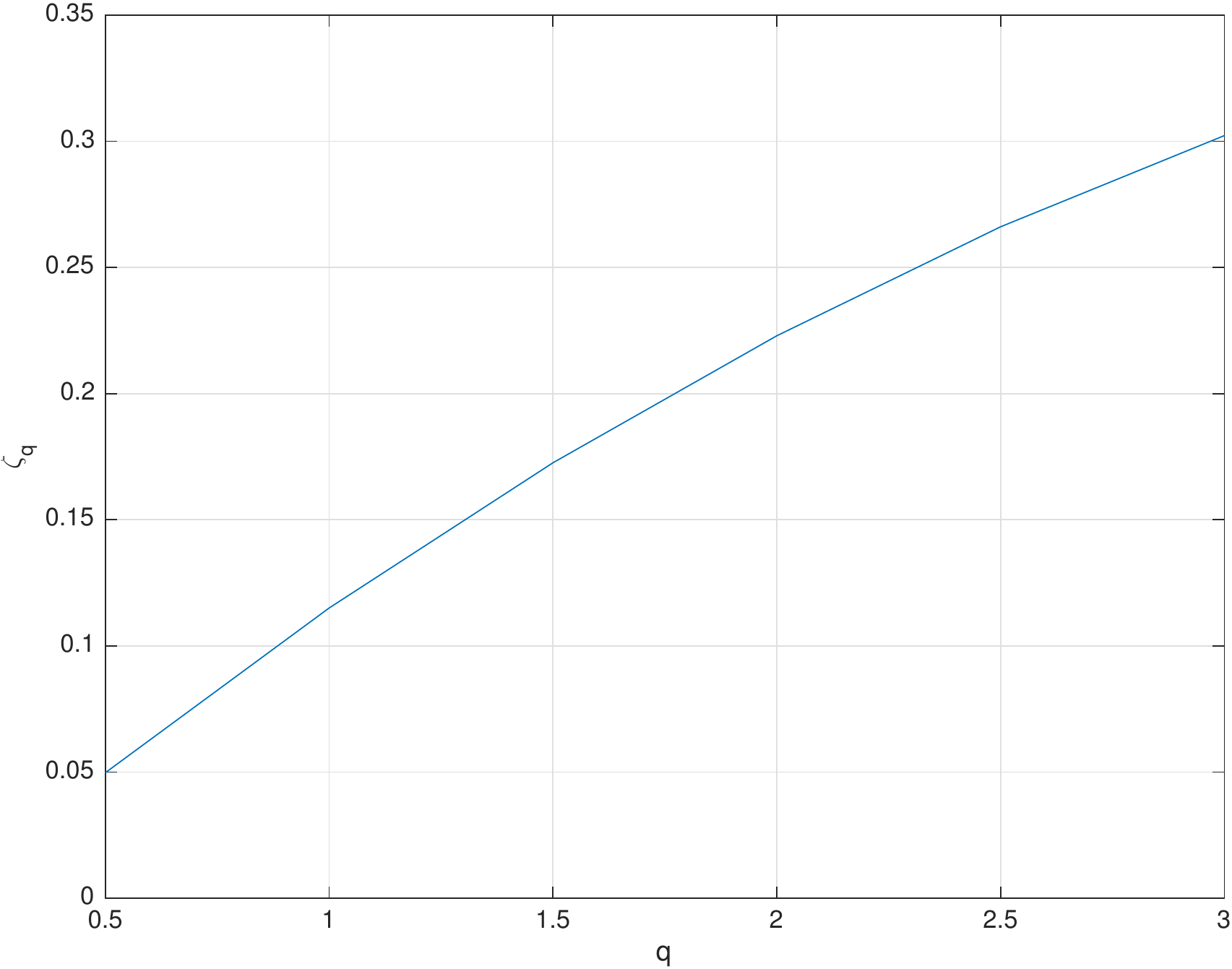}
\caption{$\log m(q,\Delta)$ vs. $\log\Delta$ for S\&P 500 realized variance estimates from OMIQF and estimating the Hurst parameter by $\zeta_q\sim Hq$. We have H = 0.10091}
\label{figLogLog2}

\
\\

\includegraphics[scale = 0.42]{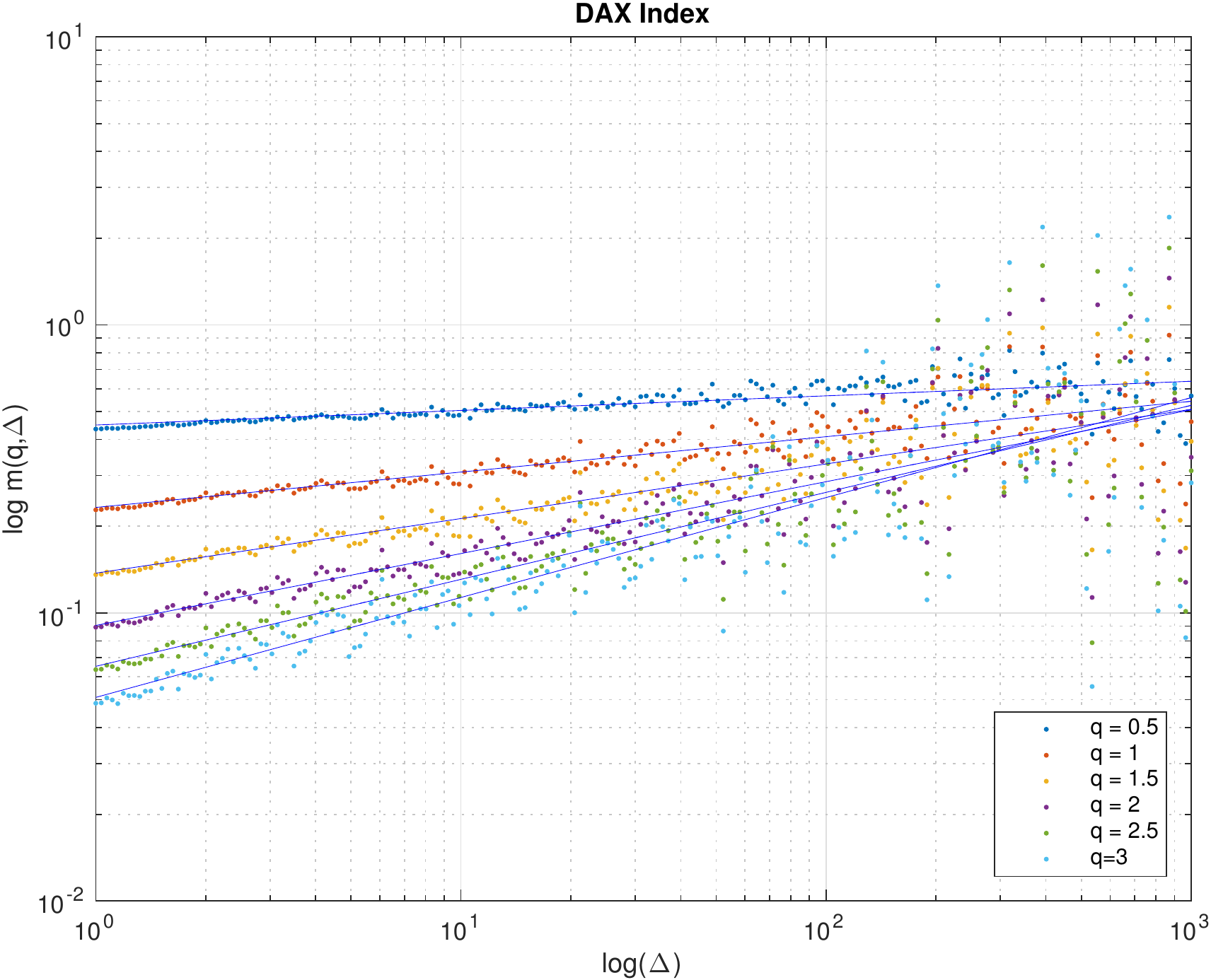}
\includegraphics[scale = 0.42]{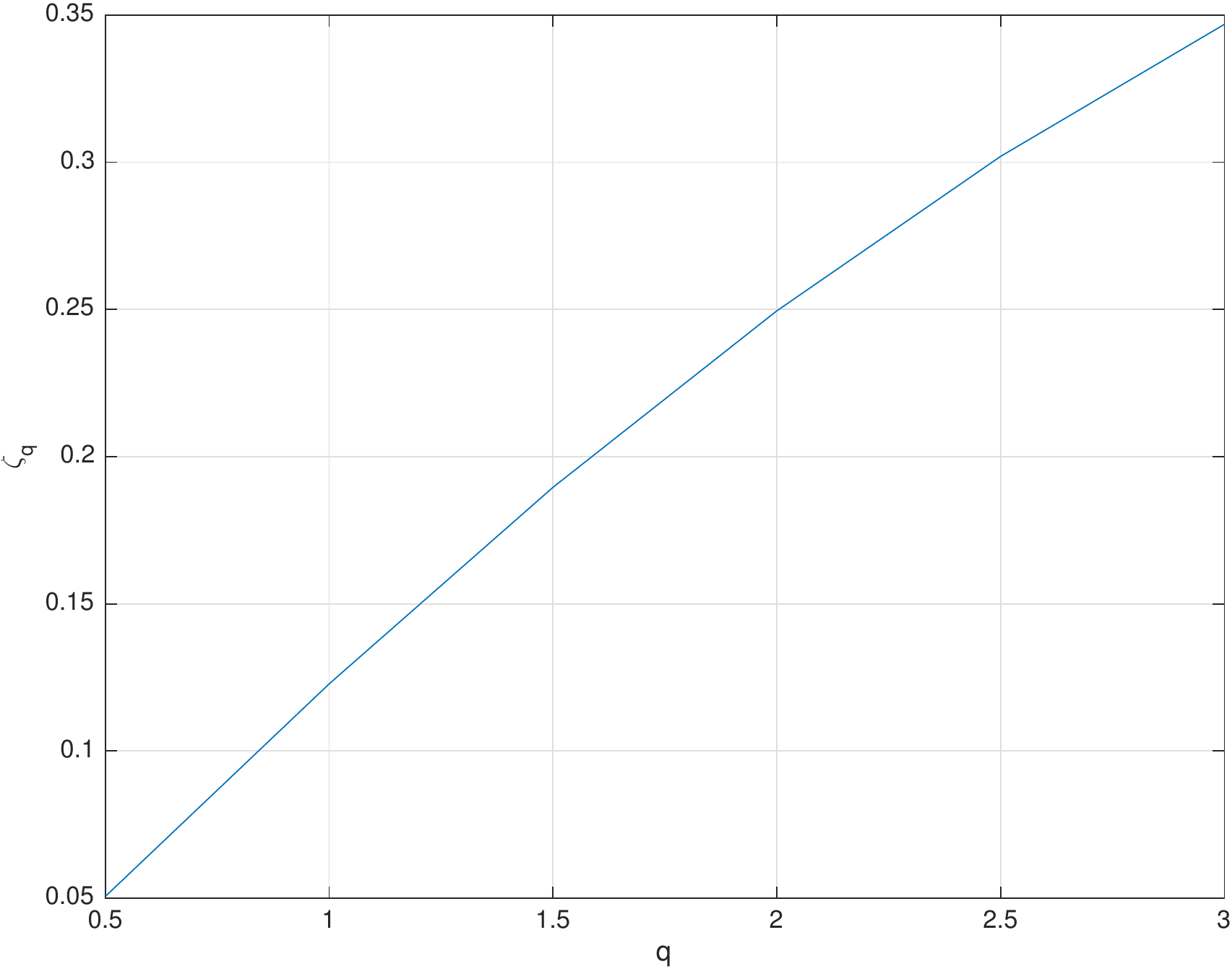}
\caption{$\log m(q,\Delta)$ as a function of $\log\Delta$ for DAX realized variance estimates from OMIQF and estimating the Hurst parameter by $\zeta_q\sim Hq$. We have H = 0.11878}
\label{figLogLog4}

\
\\

\includegraphics[scale = 0.42]{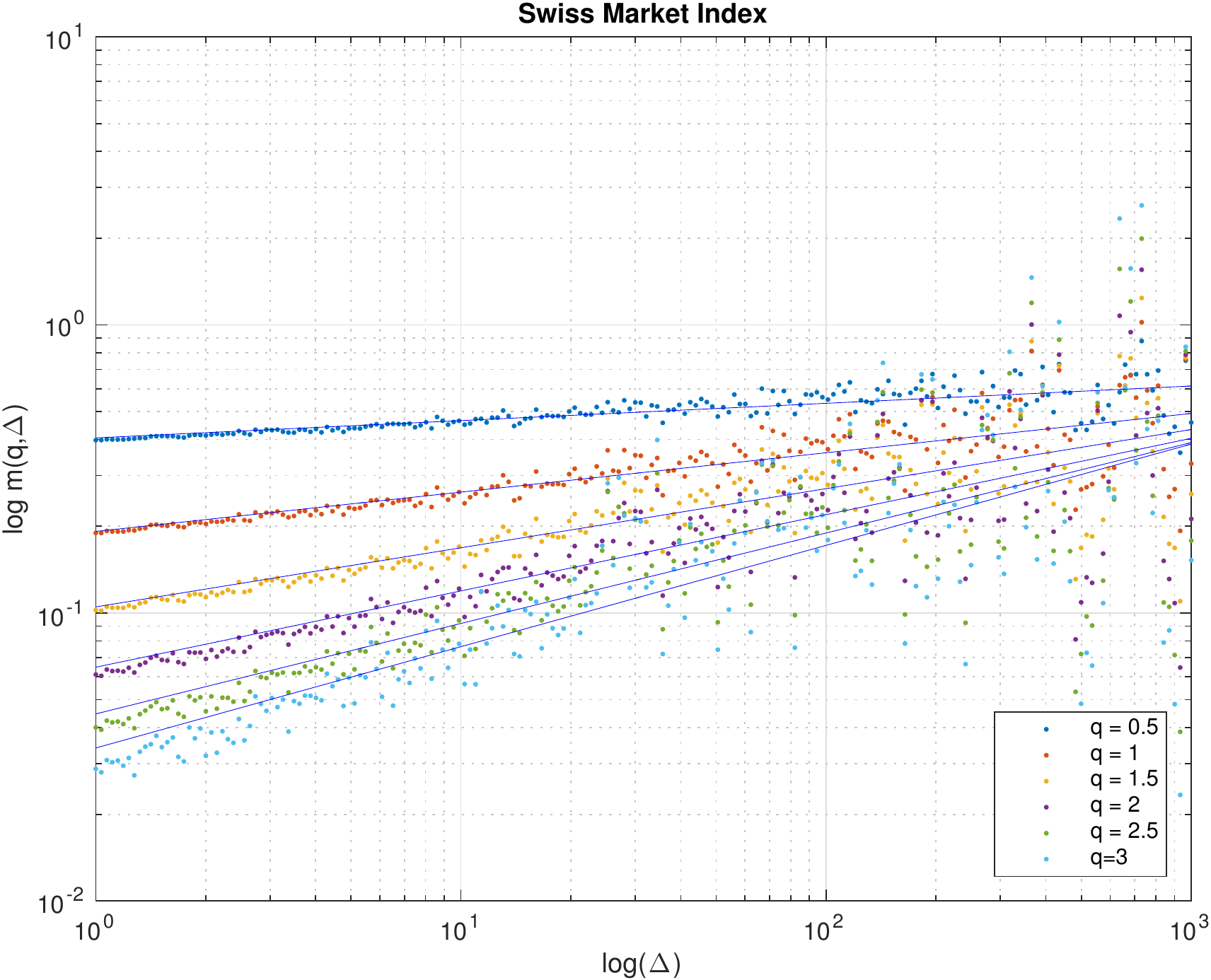}
\includegraphics[scale = 0.42]{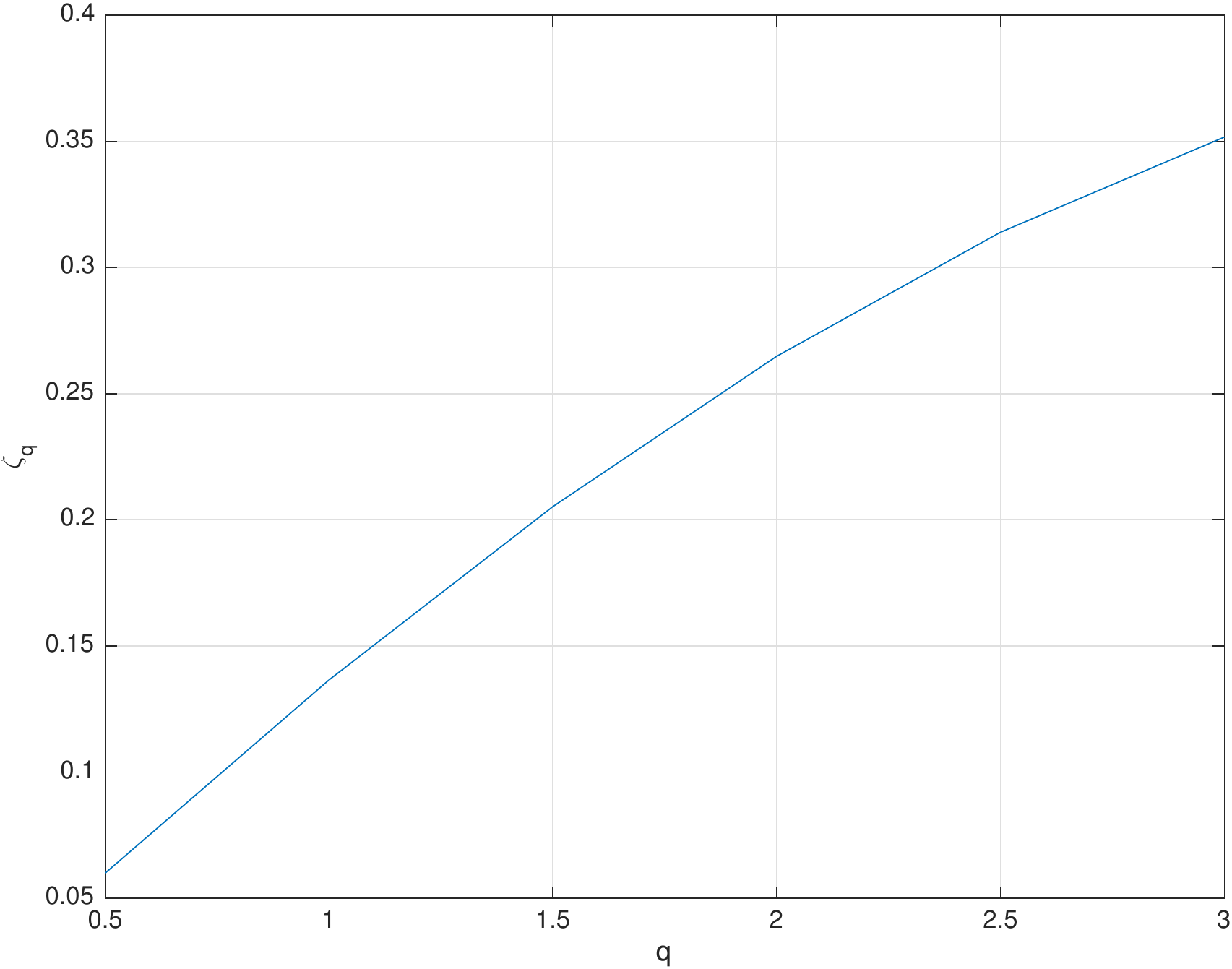}
\caption{$\log m(q,\Delta)$ vs. $\log\Delta$ for Swiss Market Index. We estimate H = 0.11714}
\label{figLogLogF}
\end{figure}

\begin{figure}
\centering
\includegraphics[scale = 0.42]{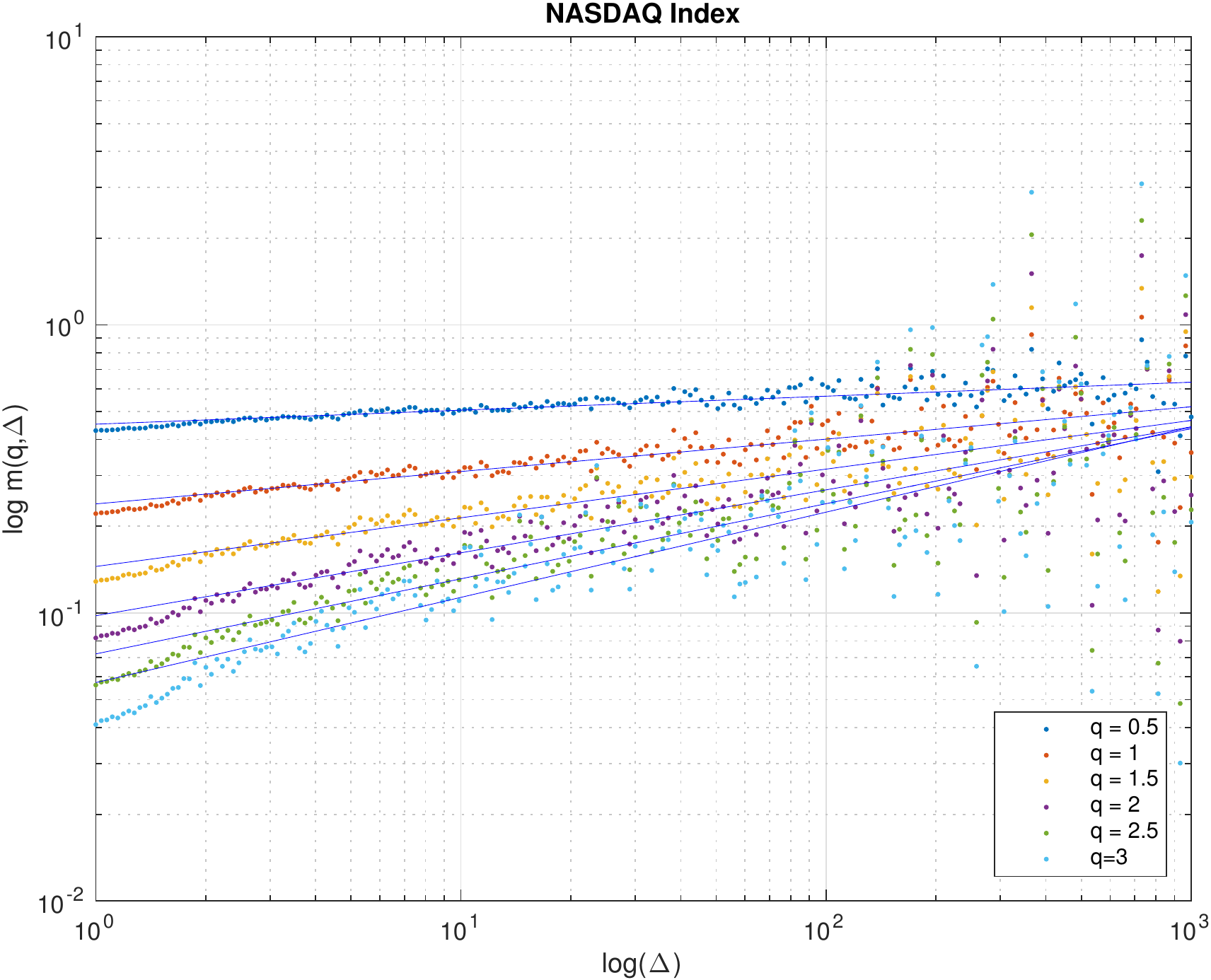}
\includegraphics[scale = 0.42]{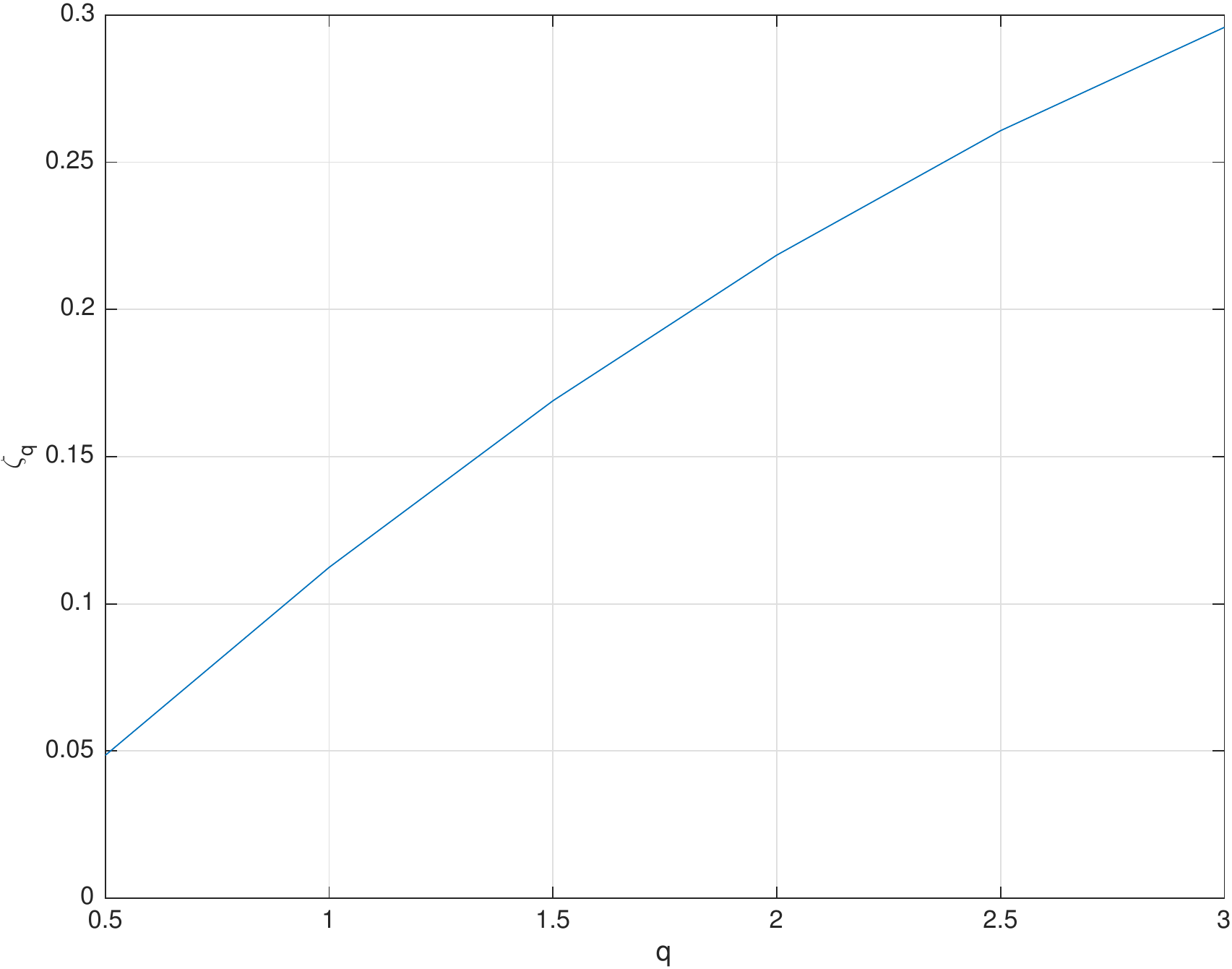}
\caption{$\log m(q,\Delta)$ vs. $\log\Delta$ for NASDAQ. We obtain H = 0.098903}
\label{figLogLog3}

\
\\

\includegraphics[scale = 0.42]{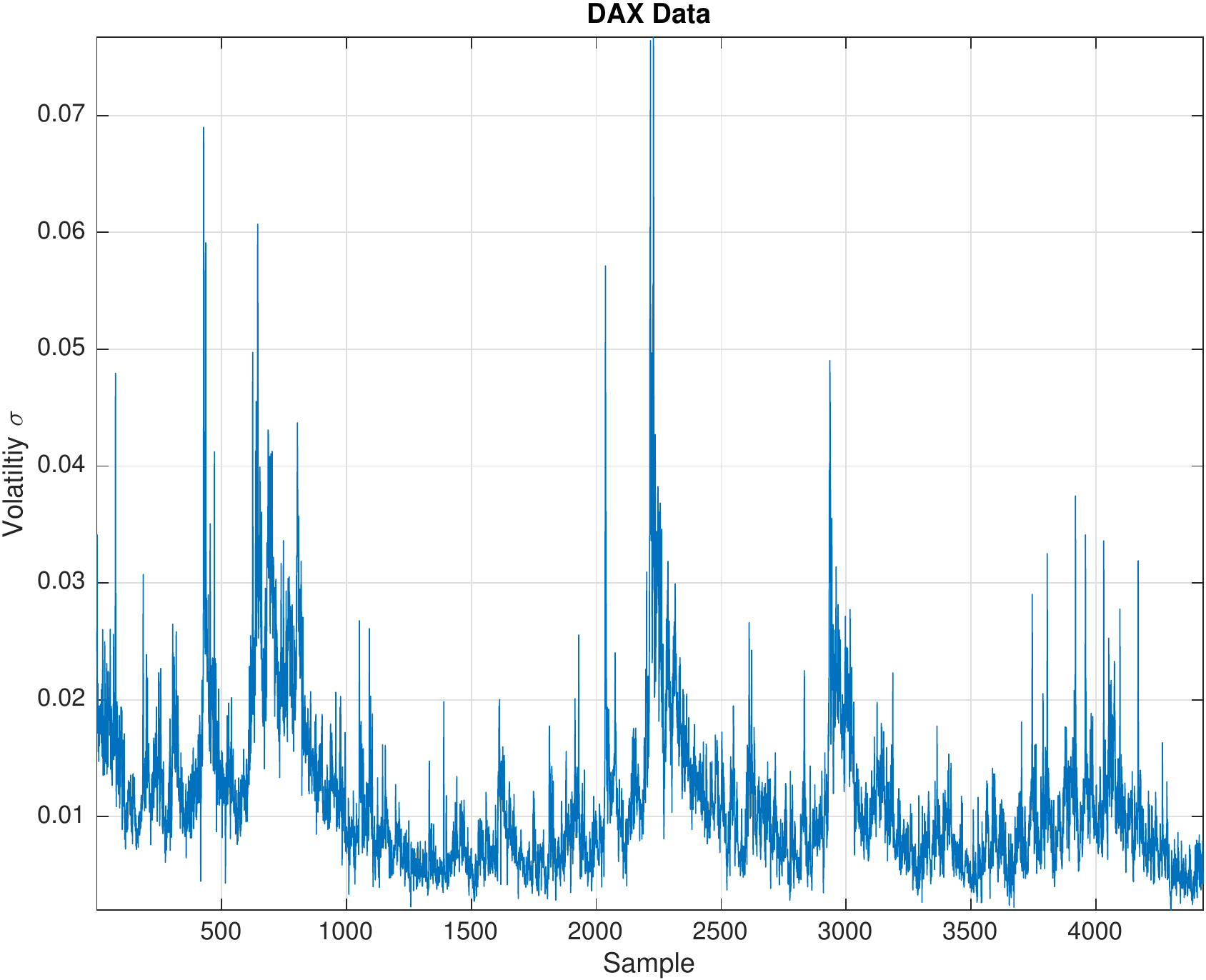}
\includegraphics[scale = 0.42]{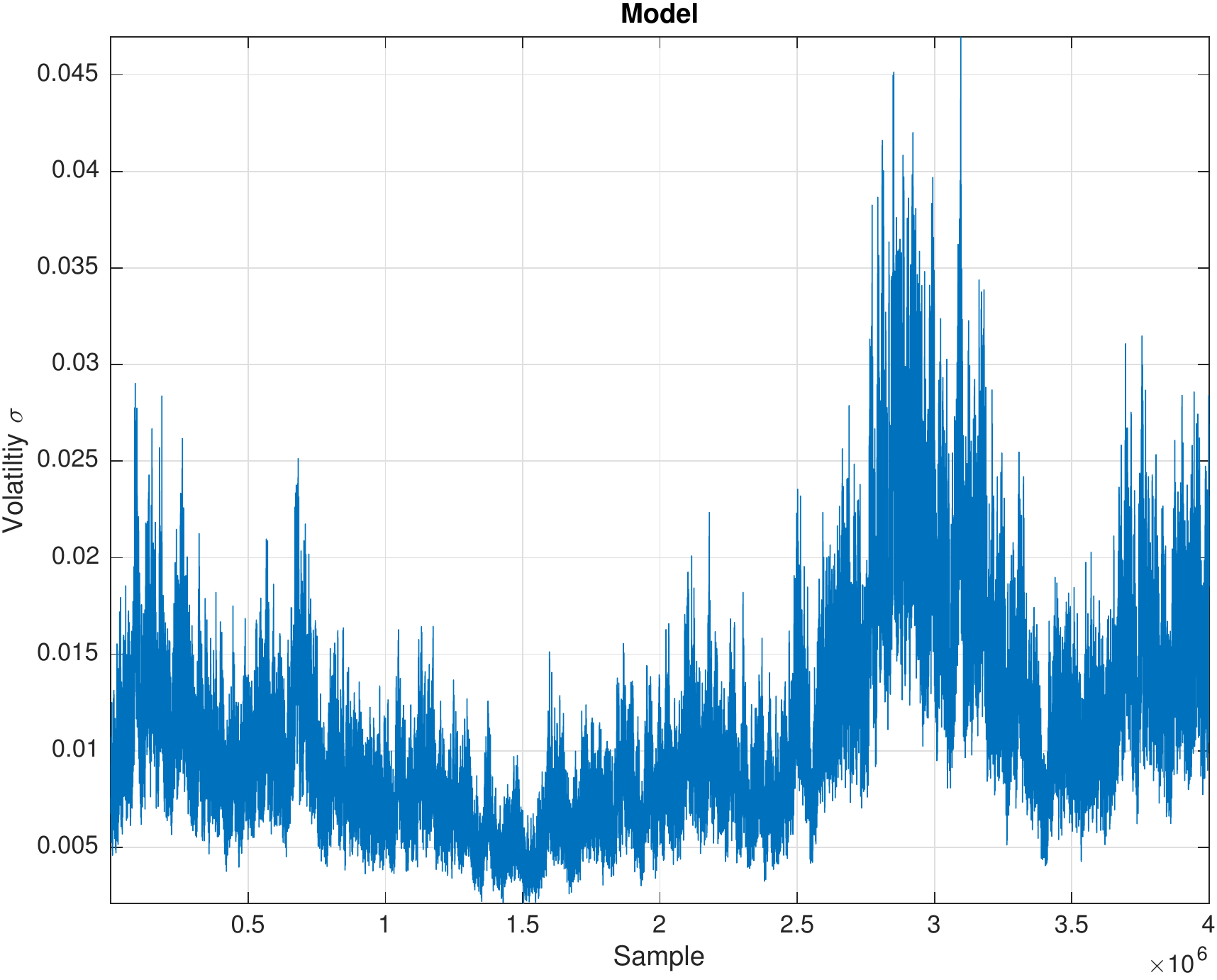}
\caption{Volatility of the DAX and of the model}
\label{DM}

\
\\

\includegraphics[scale = 0.42]{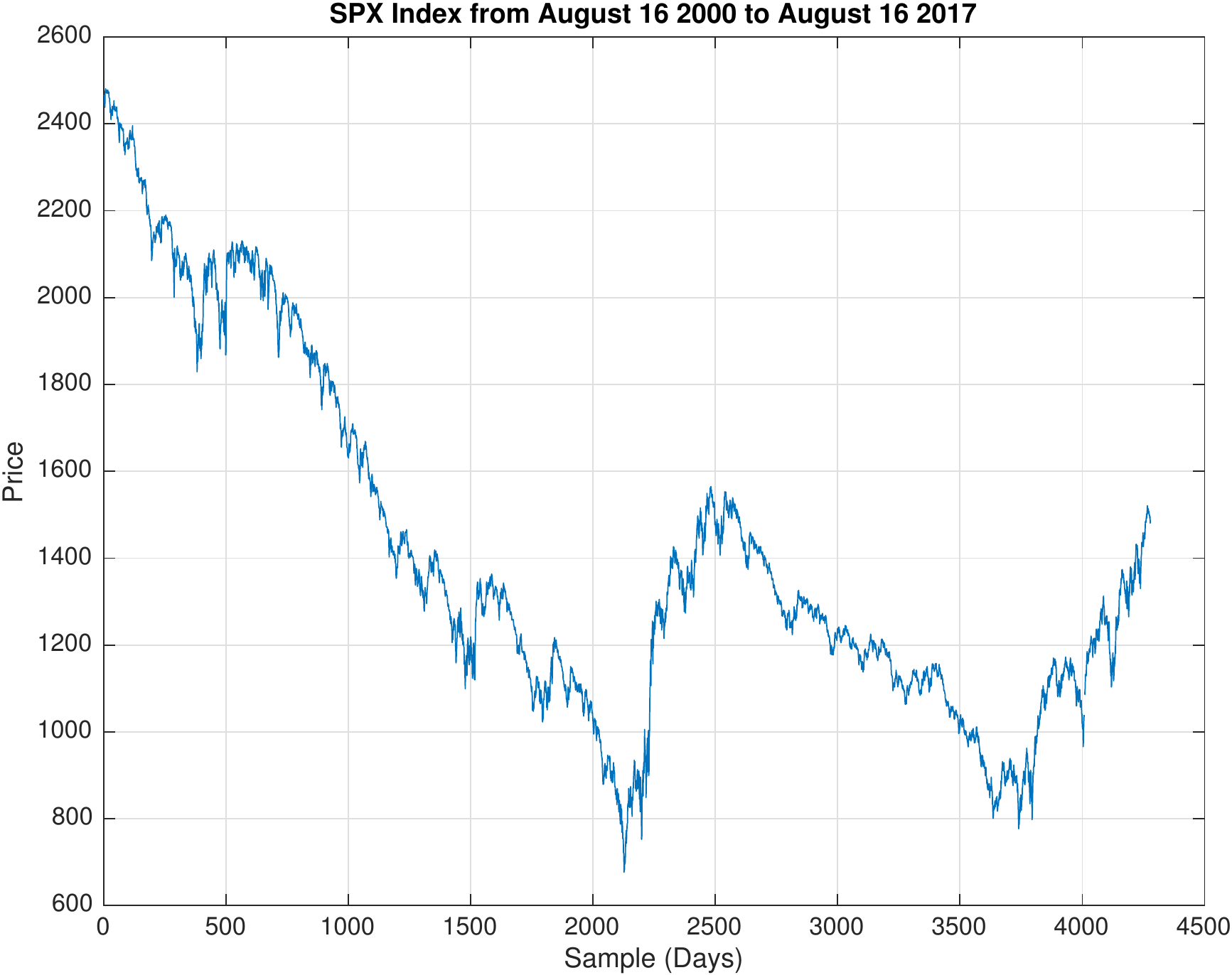}
\includegraphics[scale = 0.42]{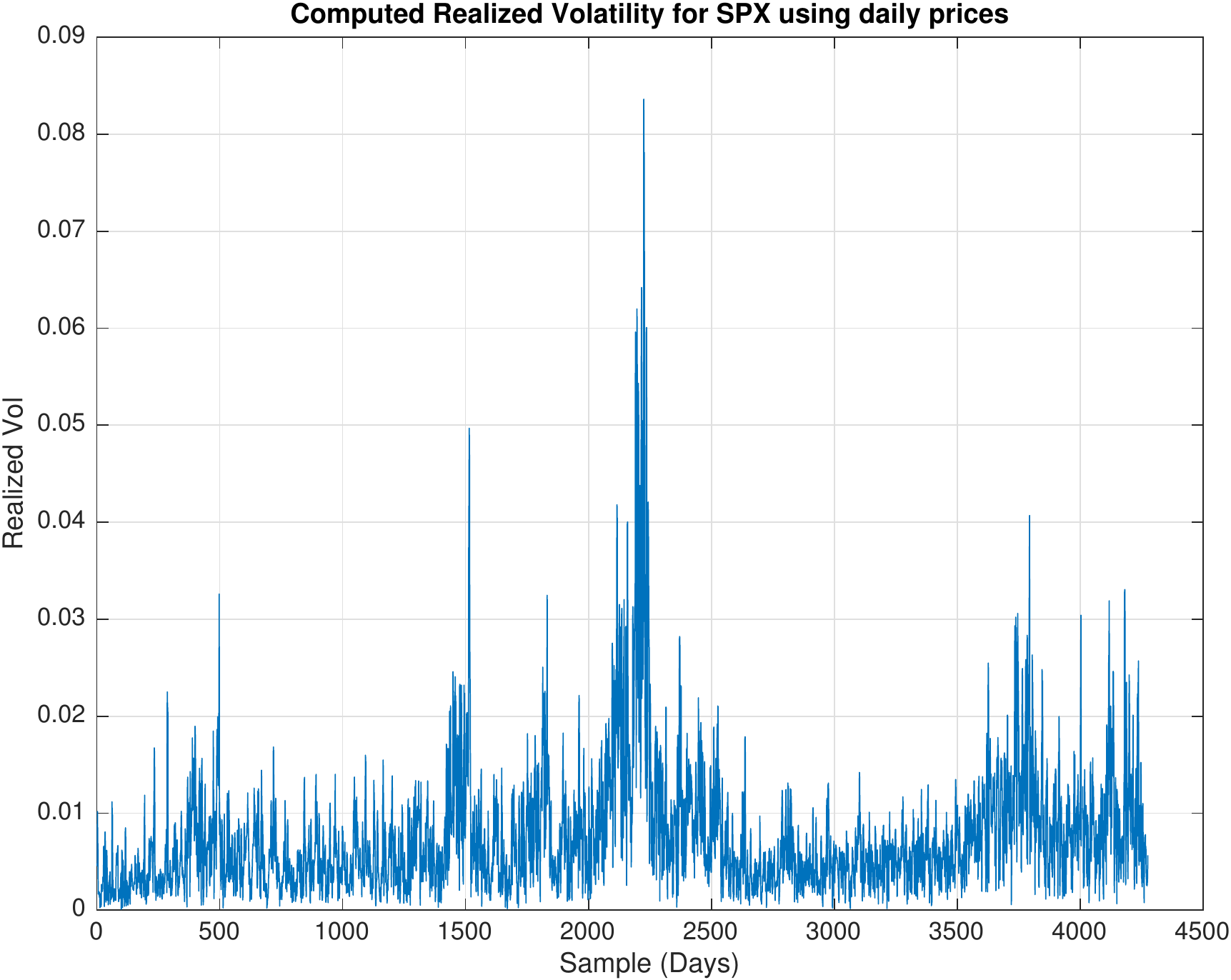}
\caption{S\&P 500 daily closing prices and calculated realized volatility. Our results match the realized volatility data from the Oxford--Man Institute for QF.}
\label{plictisit}
\end{figure}

We repeat the analysis from above for several well know equity indices and market indicators: S\&P500, DAX30, Swiss Market Index, and the NASDAQ100. We use precomputed realized variance estimates from the Oxford-Man Institute of Quantitative Finance Realized Library (OMIQF)\footnote{\href{http://realized.oxford-man.ox.ac.uk/data/download}{http://realized.oxford-man.ox.ac.uk/data/download}}. The results are given in Figures \ref{figLogLog2} to \ref{figLogLogF}. Note that we always obtain $H<0.5$ (of order $0.1$) thus confirming the results in \cite{gatheral2018volatility}. We also present in Figure \ref{DM} a sample path of the RSFV model--generated volatility together with a graph of DAX volatility. It can be noted by simple visual inspection the striking similarity between the paths. We observe that while some data points of the function $m(q,\Delta)$ for fixed $q$ may have a poorer linear fit, the main conclusion is always the same: $H<0.5$, thus confirming the \emph{rough} nature of the stochastic volatility process. 

Furthermore, we implement our own procedure for estimating the realized volatility based on daily prices, which we obtain from the Bloomberg terminal. Let $0=t_0<t_1<\cdots<t_n=T$ be a partition of the time interval $[0,T]$ into $n$ equal segments of length $\Delta t$, i.e., $t_i = \frac{iT}{n}$ for $i=0,1,\dots,n$. The majority of traded contracts define realized variance to be \cite{broadie2008pricing}: 
\[
V_d(0,n,T) = \frac{AF}{n-1} \sum_{i=0}^{n-1} \left(\ln\frac{S_{i+1}}{S_i} \right)^2
\]
for $n$ return observations, where {\it AF} is the annualized factor usually set to $AF=252$. $V_d(0,n,T)$ is called the discretely sampled realized variance from 0 to $T>0$. We implement the formula above for SPX, DAX, NASDAQ, and the Swiss Market Index. Our data consists in daily closing prices from August 16 2000 to August 16 2017. The results are plotted in Figure \ref{plictisit}.Our results are consistent with the live data for realized variance from the Oxford-Man Institute for Quantitative Finance. Further, we implemented the procedure described above for estimating the Hurst exponent using our own computed realized volatility. Once more, the results are consistent with the previous experiments.

Let $V_c(0,T)$ be the so--called continuous realized variance where one has

$$
V_c(0,T) = \lim_{n\to\infty} V_d(0,n,T).
$$
Furthermore, assuming a stochastic volatility model $\sigma_t,\ t\ge0$, it can be proved that the continuous realized variance is given by:

\[
V_c(0,T) = \frac{1}{T}\int_0^T \sigma^2_t\d t.
\]

Similar to \cite{gatheral2018volatility}, we conclude from our market analysis that the volatility is indeed {\it rough}, i.e., can be successfully modeled as an exponential of a scaled fBM. Estimation of the Hurst parameter, which is essential to the nature of fBMs, turns out to be reliable, robust, and numerically efficient.

\subsection{The lognormal fractional SABR model}

The price of underlying asset $S_t$ and its instantaneous volatility $\alpha_t$ in the lognormal fSABR model are governed by the following stochastic differential equation:
\begin{eqnarray}
&& \dfrac{dS_t}{S_t} = \alpha_t (\rho dB_t + \bar\rho dW_t), \\
&& \alpha_t = \alpha_0 e^{\nu B_t^H},
\end{eqnarray}
where $B_t$ and $W_t$ are independent Brownian motions, $\bar\rho = \sqrt{1 - \rho^2}$, and $B_t^H$ is the fractional Brownian motion driven by $B_t$. For fixed $K > 0$, we define $X_t = \log\frac{S_t}K$ and $Y_t = \alpha_t$. Then we have
\begin{eqnarray}
&& dX_t = Y_t (\rho dB_t + \bar\rho dW_t) - \frac{1}{2} Y_t^2 dt = Y_t d \tW_t - \frac{1}{2}  Y_t^2 dt, \label{eqn:fSABR-x} \\
&& Y_t = Y_0 e^{\nu B_t^H}. \label{eqn:fSABR-y}
\end{eqnarray}

In \cite{akahori2017probability} the authors derive a bridge representation for the joint transition probability density $p(t,x_t,y_t|x_0,y_0)$ of $(X_t,Y_t)$ from fSABR. Furthermore, the authors have also shown that their bridge representation can be regarded as a generalization of the McKean kernel density with respect to the Riemannian volume from $\frac{1}{y_t^2}\d x_t \d y_t$.


\subsection{Target Volatility Options Pricing via Quantum Random Numbers}

We price target volatility options (TVO) under a fBM stochastic process via MC methods using the QRNs described in the previous sections. A similar problem was approached in \cite{alos2018target}, where the authors derived two analytic approximations of TVO prices. The Python implementation using PRNs (similar to our MC pricing presented here) is available online on the GitHub platform\footnote{\href{https://github.com/studor/TVO_pricing_fSABR}{https://github.com/studor/TVO\_pricing\_fSABR}}. 

We indicate here that our method is more efficient from a numerical point of view than its classical counterpart using pseudo-random numbers generated on a classical computer. The fSABR model is used where fBM paths are generated using both QRNs and PRNs.

\begin{figure}
\centering
\includegraphics[scale = 0.51]{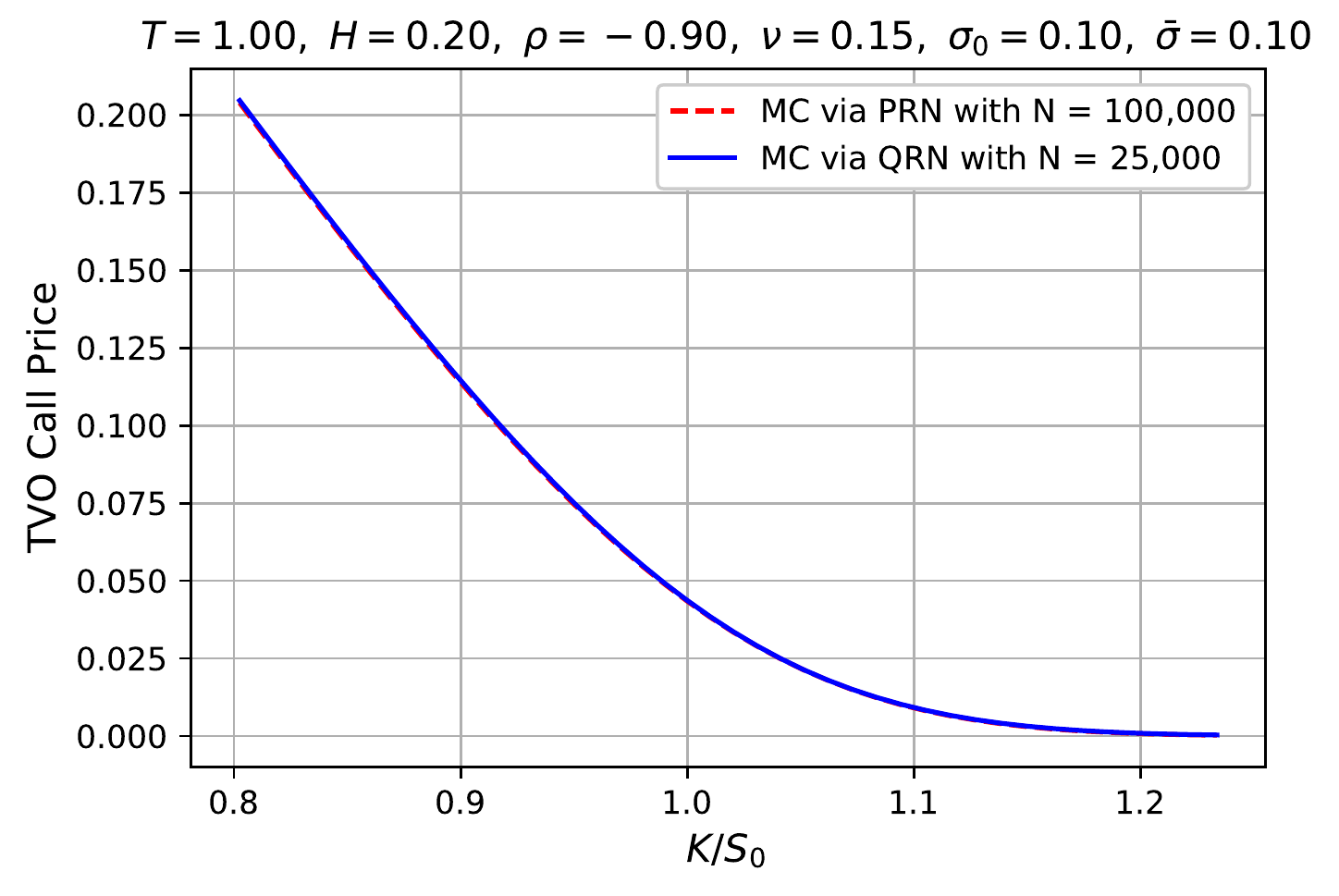}
\includegraphics[scale = 0.51]{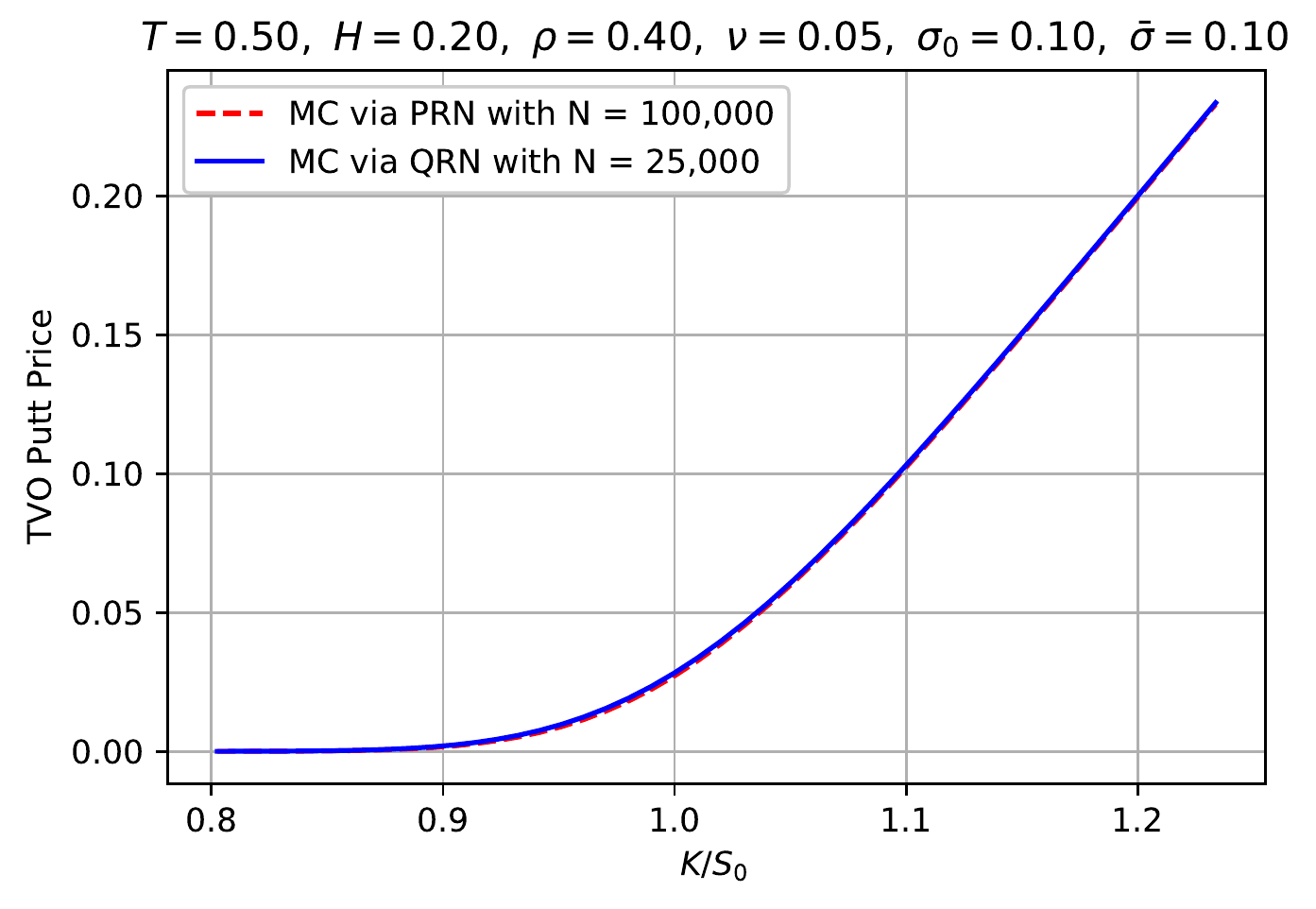}

\
\\

\includegraphics[scale = 0.51]{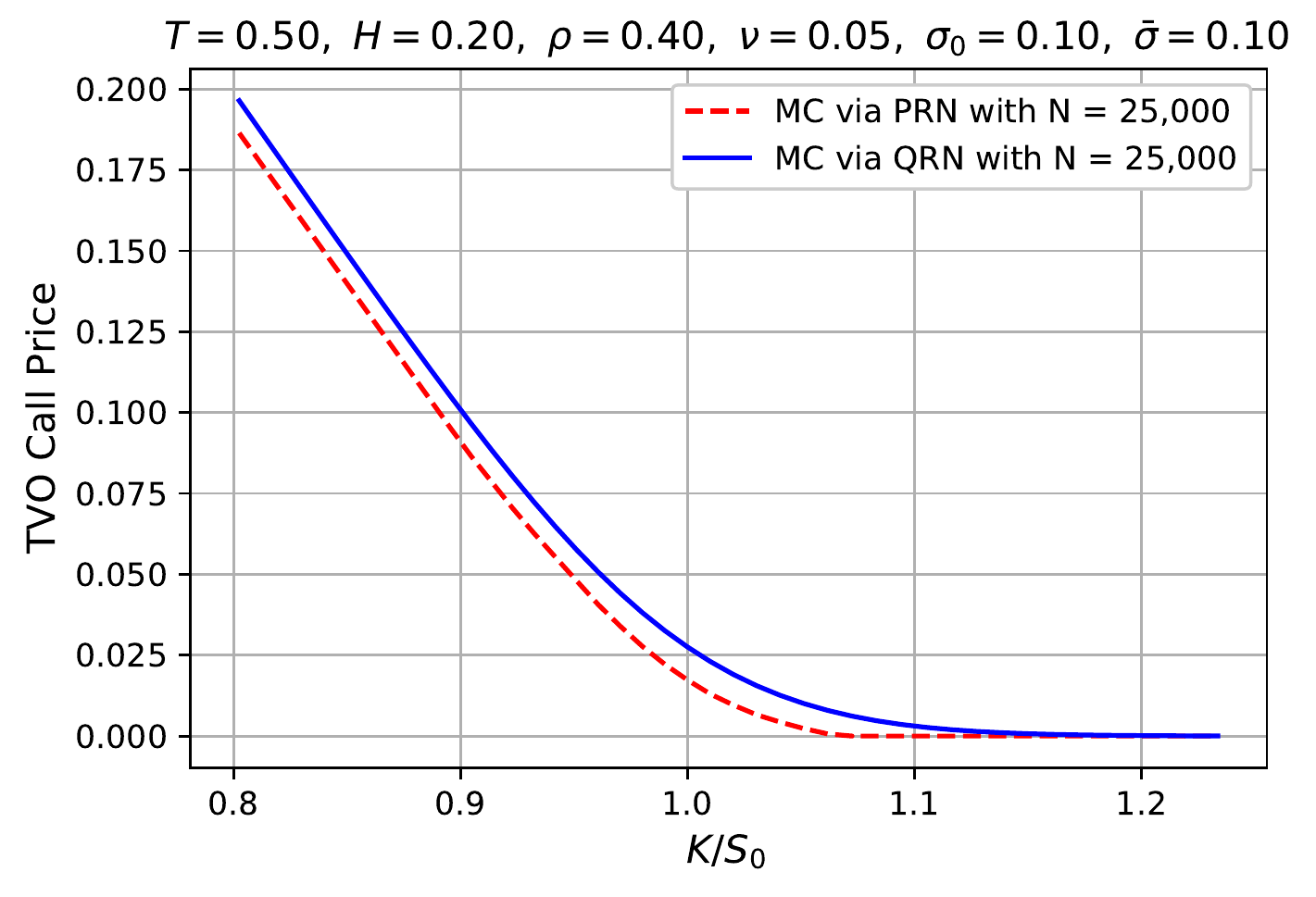}
\includegraphics[scale = 0.51]{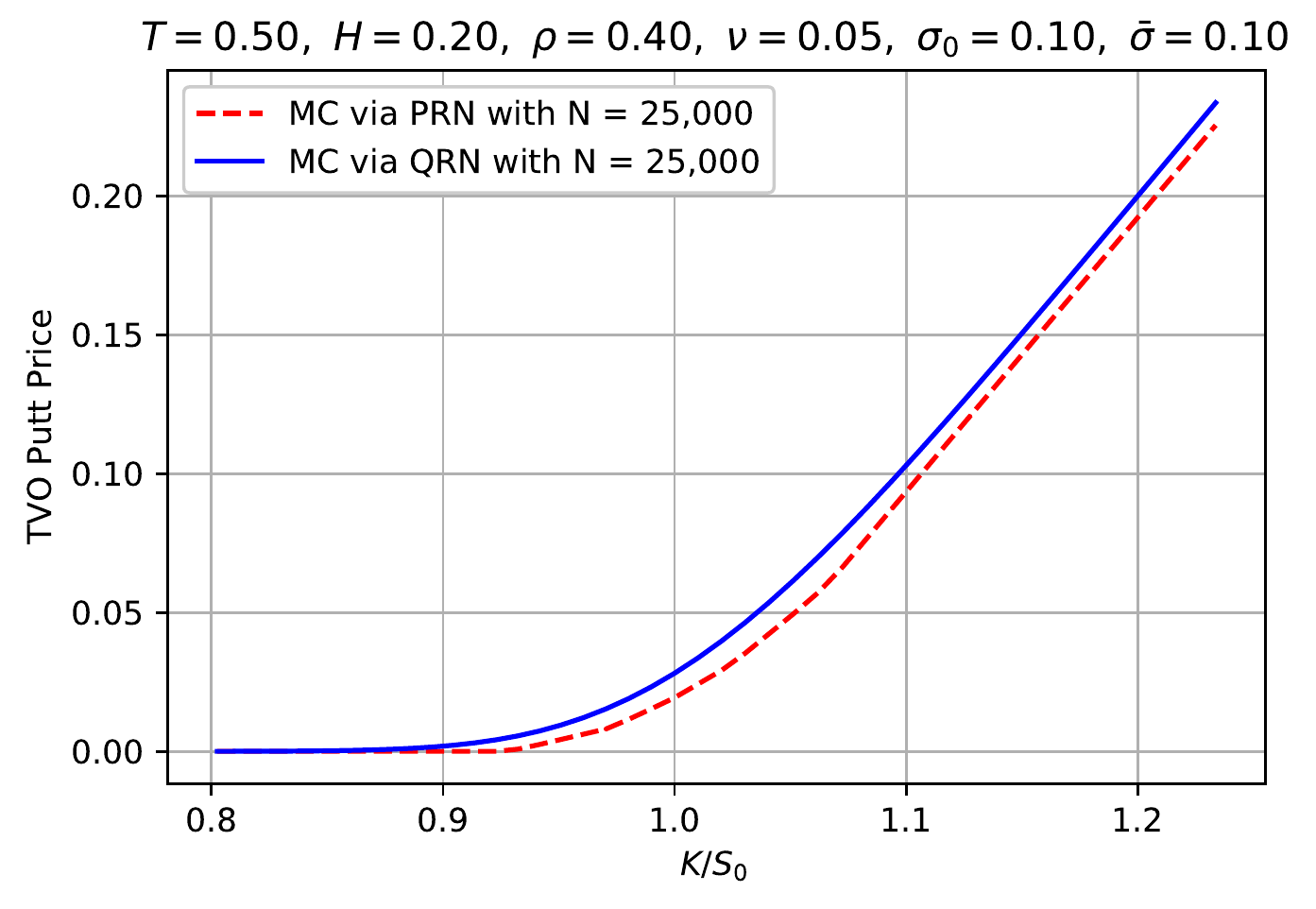}
\caption{Comparison between TVO calls/puts pricing via PRNs and QRNs. We used 44 strikes and fixed maturity $T$. It is clear that the same performance can be obtained with a smaller number of MC trials using QRNs}
\label{tvoQRN}

\
\\

\includegraphics[scale=0.5]{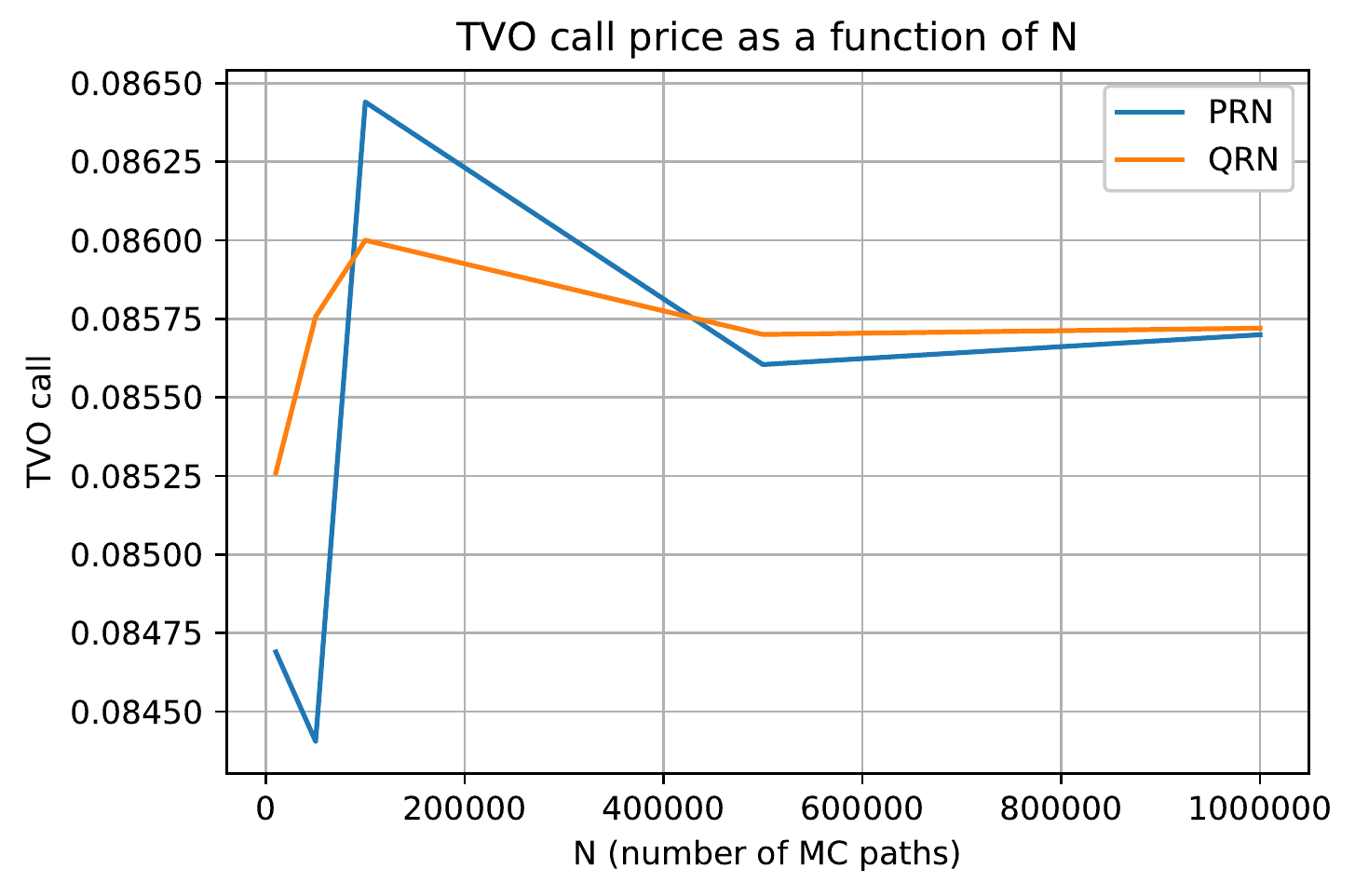}
\includegraphics[scale=0.5]{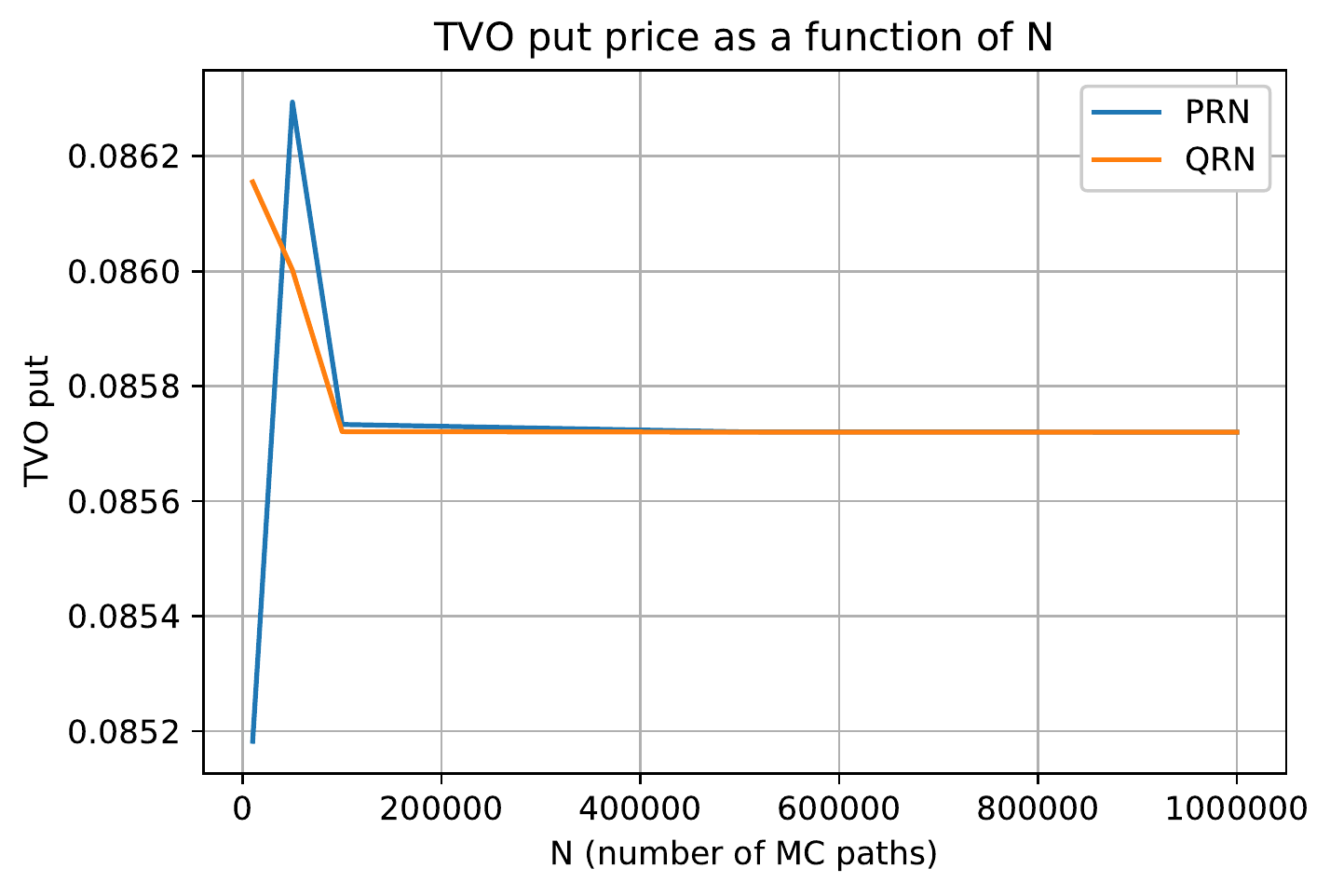}
\caption{Convergence of TVO call and put price with the number of MC paths $N$. We notice that QRN have superior properties, as the price reaches convergence faster than its classical counterpart (PRN). Parameters are $H=0.1, S_0=K=1, \sigma_0=\bar\sigma=0.3,\rho=-0.5,T=0.5,n=500.$}
\label{fig:MCconv}
\end{figure}

A {\it target volatility} (TV) call struck at $K>0$  pays off at expiry $T>0$ the amount

\begin{equation} \label{eqn:tvo-payoff}
\frac{\bar\sigma}{\sqrt{\frac1T\int_0^T \alpha_t^2 dt}} \left( S_T - K \right)^+ = \frac{K \, \bar\sigma \sqrt T}{\sqrt{\int_0^T Y_t^2 dt}} \left( e^{X_T}- 1 \right)^+,
\end{equation}
where $\bar\sigma>0$ is the (preassigned) {\it target volatility} level and $\forall x\in\R, (x)^+:=\max\{x,0\}$. Note that, if at expiry the realized volatility is higher (lower) than the target volatility, the payoff is scaled down (up) by the ratio between target volatility and realized volatility. For $t \leq T$, the price at time $t$ of a TVO call struck at $K$ with expiry $T$ is hence given by the conditional expectation under the risk neutral probability $\mathbb Q$ as 

\begin{equation}
K\, \bar\sigma \sqrt T \, \Eof{\left. \frac1{\sqrt{\int_0^T Y_\tau^2 d\tau}} \left( e^{X_T}- 1 \right)^+\right|\cF_t} \label{eqn:tvo-price-t}
\end{equation} 
provided the expectation is finite. Finally, we price both TVO call options and the following new contract type called a {\it target volatility put option}, having the payoff:

\bel{tvoPut}
\frac{1}{\bar\sigma}\sqrt{\frac{1}{T} \int_0^T \alpha_t^2\d t}\ (K-S_T)^+ = \frac{K}{\bar\sigma\sqrt T} \sqrt{ \int_0^T Y_t^2\d t}\ (1-e^{X_T})^+.
\ee
Note that the target volatility is in the denominator instead of the numerator, see \cite{alos2018target} for comparison. MC methods via PRNs and QRNs were implemented and the results are shown in Figures \ref{tvoQRN} and \ref{fig:MCconv}. Note that once again, we observe more accurate pricing using QRN generated MC paths. Specifically, Figure \ref{tvoQRN} compares the convergence a MC simulation based on QRNs using 25k paths versus MC PRNs using 100k paths. Figure \ref{fig:MCconv} shows faster convergence of MC QRN as a function of the number of MC paths $N$. We conclude that by using QRNs, MC simulations for TVO pricing become far more efficient than its classical counterpart.

\section{Conclusions and future research}

In this paper, we showed a novel use of quantum random numbers for simulating stochastic processes. The photonic quantum system used here produced truly random numbers that passed all standard randomness tests such as NIST and Dieharder. Furthermore, the efficiency, accuracy, and execution time of our QRNs were demonstrated by the rapid convergence and robustness of using these numbers to simulate fractional Brownian motion -- a highly nontrivial stochastic process being both non-Markovian while maintaining strict scaling properties. Finally, we successfully priced target volatility options under the fractional SABR model, a relevant and interesting application for both academics and quantitative finance practitioners.


The results of this paper can be easily generalized: we conjecture that QRN MC trials for path dependent-exotic options and multi-asset correlated options will turn out to be computationally less expensive and more accurate. Some future applications include: (i) multi-dimensional correlated simulation paths following arbitrary distributions may be efficiently created via the quantum methods described in this paper; (ii) the paths generated in (i) are crucial for pricing both multi-asset derivatives but more importantly, for use in the risk analysis for large portfolios of such options \cite{chatterjee}; (iii) enterprise wide regulatory risk calculations (Basel III, CVA, XVA, etc., see \cite{chatterjee}) require and enormous amounts of multi-dimensional correlated simulation paths following arbitrary distributions. QRN will show their true strength under such pricing scenarios which is a future research.

\newpage
\bibliographystyle{plain}
\bibliography{referencesQual}

\end{document}

%% file: notation.tex
\usepackage{amsfonts}
\usepackage{epsfig}
\usepackage{graphicx}
\usepackage{color}
\usepackage{fancyhdr}
\usepackage{url}
\usepackage{epstopdf}
\usepackage{dsfont}
\usepackage{amssymb,amsmath}
\usepackage{amsfonts}
\usepackage{enumerate}
\usepackage{bigints}
\usepackage{lipsum}
\usepackage{algorithmic}

\newcommand{\R}{\mathbb{R}} 

\newcommand{\Z}{\mathbb{Z}}

\newcommand{\x}{\times }

\newcommand{\ba}{\left[ \begin{array}}
\newcommand{\baa}{\begin{array}}
\newcommand{\ea}{\end{array} \right]}
\newcommand{\eaa}{\end{array}}

\newcommand{\be}{\begin{equation}}
\newcommand{\ee}{\end{equation}}
\newcommand{\bel}{\begin{equation}\label}

\newcommand{\ov}{\overline}
\renewcommand{\tilde}{\widetilde}

\renewcommand{\bar}{\ov}

\renewcommand{\d}{{\mathrm{d}}}

\newcommand{\CC}{{{\mbox{\rm \hspace*{0.05ex}
\rule[.18ex]{.18ex}{1.24ex} \kern -.65em C}}}}

\newcommand{\E}{\mathbb{E}}
\newcommand{\Eof}[1]{\mathbb{E}\left[ #1 \right]}

\newcommand{\beaa}{\begin{eqnarray*}}
\newcommand{\eeaa}{\end{eqnarray*}}
\newcommand{\bea}{\begin{eqnarray}}
\newcommand{\eea}{\end{eqnarray}}
\newcommand{\ben}{\begin{enumerate}}
\newcommand{\een}{\end{enumerate}}
\newcommand{\bit}{\begin{itemize}}
\newcommand{\eit}{\end{itemize}}

\newcommand{\tW}{\tilde{W}}

\newcommand{\cF}{\mathcal{F}}